\documentstyle[10pt,epsf,epsfig,dp_delphititle]{dp_delphi}
\makeindex
\def\DpPaperGroup{PH-EP}
\def\DpPaperRef{2007-036}
\def\DpDate{6 October 2007}
\def\DpAuthors{DELPHI Collaboration}
\def\DpSubmit{(Accepted by Eur. Phys. J. C)}
\def\DpTitle{{
Study of W boson polarisations and 
Triple Gauge boson Couplings
in the reaction \boldmath $e^+ e^- \rightarrow W^+ W^-$ at 
LEP 2
}}
\def\DpComment{}
\def\DpEMail{}

\begin{document}
\makeatletter
 
\makeatother
\begin{titlepage}
\pagenumbering{roman}
\CERNpreprint{\DpPaperGroup}{\DpPaperRef} 
\date{{\small\DpDate}} 
\title{\DpTitle} 
\address{\DpAuthors} 

\begin{shortabs} 
\noindent
\noindent

A determination of the single W Spin Density Matrix (SDM) elements 
in the reaction $e^+ e^- \rightarrow W^+ W^- \rightarrow l \nu q \bar{q} (l=e/\mu)$ 
is reported at centre-of-mass energies between 189 and 209 GeV. 
The  
data sample used corresponds to 
an integrated luminosity of 520 $\rm pb^{-1}$ taken by DELPHI 
between 1998 and 2000. 
\\
The single W SDM elements, 
$\rho_{\tau\tau'}^{W^{\pm}}$ ($\tau,\tau' = \pm$ 1 or 0), are determined 
as a function of the $\rm W^-$ production angle with respect to the $\rm e^-$ beam direction and
are obtained from measurements of the 
 W decay products by the application of suitable projection operators, $\Lambda_{\tau\tau'}$, which
assume the V-A coupling of the W boson to fermions.\\
The measured SDM elements  are  used to obtain the fraction of longitudinally polarised Ws, with the result:
\begin{eqnarray*}
\frac{\sigma_L}{\sigma_{tot}} = 24.9 \pm 4.5(stat) \pm 2.2(syst)
\%
\end{eqnarray*}
at a mean energy of 198 GeV. 
The SDM elements are also used to determine  the 
Triple Gauge Couplings $ \Delta g_1^Z, \Delta \kappa_{\gamma}, \lambda_{\gamma}$ and 
$g_4^Z, \tilde{\kappa}_Z$ and $\tilde{\lambda}_Z$. 
For the CP-violating couplings the results of single parameter fits are:
\begin{eqnarray*}
g_4^Z             & = &-0.39^{+0.19}_{-0.20} \\
\tilde{\kappa}_Z  & = &-0.09^{+0.08}_{-0.05} \\
\tilde{\lambda}_Z & = &-0.08\pm0.07 .
\end{eqnarray*}
The errors are a combination of statistical and systematic errors.
All results are consistent with the Standard Model.


\end{shortabs}

\vfill

\begin{center}
\DpSubmit \ \\ 
\DpComment \ \\
\DpEMail \ \\
\end{center}
\vfill
\clearpage

\headsep 10.0pt

\addtolength{\textheight}{10mm}
\addtolength{\footskip}{-5mm}
\begingroup
%
\newcommand{\DpName}[2]{\hbox{#1$^{\ref{#2}}$},\hfill}
\newcommand{\DpNameTwo}[3]{\hbox{#1$^{\ref{#2},\ref{#3}}$},\hfill}
\newcommand{\DpNameThree}[4]{\hbox{#1$^{\ref{#2},\ref{#3},\ref{#4}}$},\hfill}
\newskip\Bigfill \Bigfill = 0pt plus 1000fill
\newcommand{\DpNameLast}[2]{\hbox{#1$^{\ref{#2}}$}\hspace{\Bigfill}}
%
\footnotesize
\noindent
\DpName{J.Abdallah}{LPNHE}
\DpName{P.Abreu}{LIP}
\DpName{W.Adam}{VIENNA}
\DpName{P.Adzic}{DEMOKRITOS}
\DpName{T.Albrecht}{KARLSRUHE}
\DpName{R.Alemany-Fernandez}{CERN}
\DpName{T.Allmendinger}{KARLSRUHE}
\DpName{P.P.Allport}{LIVERPOOL}
\DpName{U.Amaldi}{MILANO2}
\DpName{N.Amapane}{TORINO}
% ...
\normalsize
\endgroup
\newpage

\titlefoot{Department of Physics and Astronomy, Iowa State
     University, Ames IA 50011-3160, USA
    \label{AMES}}
\titlefoot{IIHE, ULB-VUB,
     Pleinlaan 2, B-1050 Brussels, Belgium
    \label{BRUSSELS}}
\titlefoot{Physics Laboratory, University of Athens, Solonos Str.
     104, GR-10680 Athens, Greece
    \label{ATHENS}}
\titlefoot{Department of Physics, University of Bergen,
     All\'egaten 55, NO-5007 Bergen, Norway
    \label{BERGEN}}
\titlefoot{Dipartimento di Fisica, Universit\`a di Bologna and INFN,
     Via Irnerio 46, IT-40126 Bologna, Italy
    \label{BOLOGNA}}
\titlefoot{Centro Brasileiro de Pesquisas F\'{\i}sicas, rua Xavier Sigaud 150,
     BR-22290 Rio de Janeiro, Brazil
    \label{BRASIL-CBPF}}
\titlefoot{Inst. de F\'{\i}sica, Univ. Estadual do Rio de Janeiro,
     rua S\~{a}o Francisco Xavier 524, Rio de Janeiro, Brazil
    \label{BRASIL-IFUERJ}}
\titlefoot{Coll\`ege de France, Lab. de Physique Corpusculaire, IN2P3-CNRS,
     FR-75231 Paris Cedex 05, France
    \label{CDF}}
\titlefoot{CERN, CH-1211 Geneva 23, Switzerland
    \label{CERN}}
\titlefoot{Institut de Recherches Subatomiques, IN2P3 - CNRS/ULP - BP20,
     FR-67037 Strasbourg Cedex, France
    \label{CRN}}
\titlefoot{Now at DESY-Zeuthen, Platanenallee 6, D-15735 Zeuthen, Germany
    \label{DESY}}
\titlefoot{Institute of Nuclear Physics, N.C.S.R. Demokritos,
     P.O. Box 60228, GR-15310 Athens, Greece
    \label{DEMOKRITOS}}
\titlefoot{FZU, Inst. of Phys. of the C.A.S. High Energy Physics Division,
     Na Slovance 2, CZ-182 21, Praha 8, Czech Republic
    \label{FZU}}
\titlefoot{Dipartimento di Fisica, Universit\`a di Genova and INFN,
     Via Dodecaneso 33, IT-16146 Genova, Italy
    \label{GENOVA}}
\titlefoot{Institut des Sciences Nucl\'eaires, IN2P3-CNRS, Universit\'e
     de Grenoble 1, FR-38026 Grenoble Cedex, France
    \label{GRENOBLE}}
\titlefoot{Helsinki Institute of Physics and Department of Physical Sciences,
     P.O. Box 64, FIN-00014 University of Helsinki, 
     \indent~~Finland
    \label{HELSINKI}}
\titlefoot{Joint Institute for Nuclear Research, Dubna, Head Post
     Office, P.O. Box 79, RU-101 000 Moscow, Russian Federation
    \label{JINR}}
\titlefoot{Institut f\"ur Experimentelle Kernphysik,
     Universit\"at Karlsruhe, Postfach 6980, DE-76128 Karlsruhe,
     Germany
    \label{KARLSRUHE}}
\titlefoot{Institute of Nuclear Physics PAN,Ul. Radzikowskiego 152,
     PL-31142 Krakow, Poland
    \label{KRAKOW1}}
\titlefoot{Faculty of Physics and Nuclear Techniques, University of Mining
     and Metallurgy, PL-30055 Krakow, Poland
    \label{KRAKOW2}}
\titlefoot{Universit\'e de Paris-Sud, Lab. de l'Acc\'el\'erateur
     Lin\'eaire, IN2P3-CNRS, B\^{a}t. 200, FR-91405 Orsay Cedex, France
    \label{LAL}}
\titlefoot{School of Physics and Chemistry, University of Lancaster,
     Lancaster LA1 4YB, UK
    \label{LANCASTER}}
\titlefoot{LIP, IST, FCUL - Av. Elias Garcia, 14-$1^{o}$,
     PT-1000 Lisboa Codex, Portugal
    \label{LIP}}
\titlefoot{Department of Physics, University of Liverpool, P.O.
     Box 147, Liverpool L69 3BX, UK
    \label{LIVERPOOL}}
\titlefoot{Dept. of Physics and Astronomy, Kelvin Building,
     University of Glasgow, Glasgow G12 8QQ, UK
    \label{GLASGOW}}
\titlefoot{LPNHE, IN2P3-CNRS, Univ.~Paris VI et VII, Tour 33 (RdC),
     4 place Jussieu, FR-75252 Paris Cedex 05, France
    \label{LPNHE}}
\titlefoot{Department of Physics, University of Lund,
     S\"olvegatan 14, SE-223 63 Lund, Sweden
    \label{LUND}}
\titlefoot{Universit\'e Claude Bernard de Lyon, IPNL, IN2P3-CNRS,
     FR-69622 Villeurbanne Cedex, France
    \label{LYON}}
\titlefoot{Dipartimento di Fisica, Universit\`a di Milano and INFN-MILANO,
     Via Celoria 16, IT-20133 Milan, Italy
    \label{MILANO}}
\titlefoot{Dipartimento di Fisica, Univ. di Milano-Bicocca and
     INFN-MILANO, Piazza della Scienza 3, IT-20126 Milan, Italy
    \label{MILANO2}}
\titlefoot{IPNP of MFF, Charles Univ., Areal MFF,
     V Holesovickach 2, CZ-180 00, Praha 8, Czech Republic
    \label{NC}}
\titlefoot{NIKHEF, Postbus 41882, NL-1009 DB
     Amsterdam, The Netherlands
    \label{NIKHEF}}
\titlefoot{National Technical University, Physics Department,
     Zografou Campus, GR-15773 Athens, Greece
    \label{NTU-ATHENS}}
\titlefoot{Physics Department, University of Oslo, Blindern,
     NO-0316 Oslo, Norway
    \label{OSLO}}
\titlefoot{Dpto. Fisica, Univ. Oviedo, Avda. Calvo Sotelo
     s/n, ES-33007 Oviedo, Spain
    \label{OVIEDO}}
\titlefoot{Department of Physics, University of Oxford,
     Keble Road, Oxford OX1 3RH, UK
    \label{OXFORD}}
\titlefoot{Dipartimento di Fisica, Universit\`a di Padova and
     INFN, Via Marzolo 8, IT-35131 Padua, Italy
    \label{PADOVA}}
\titlefoot{Rutherford Appleton Laboratory, Chilton, Didcot
     OX11 OQX, UK
    \label{RAL}}
\titlefoot{Dipartimento di Fisica, Universit\`a di Roma II and
     INFN, Tor Vergata, IT-00173 Rome, Italy
    \label{ROMA2}}
\titlefoot{Dipartimento di Fisica, Universit\`a di Roma III and
     INFN, Via della Vasca Navale 84, IT-00146 Rome, Italy
    \label{ROMA3}}
\titlefoot{DAPNIA/Service de Physique des Particules,
     CEA-Saclay, FR-91191 Gif-sur-Yvette Cedex, France
    \label{SACLAY}}
\titlefoot{Instituto de Fisica de Cantabria (CSIC-UC), Avda.
     los Castros s/n, ES-39006 Santander, Spain
    \label{SANTANDER}}
\titlefoot{Inst. for High Energy Physics, Serpukov
     P.O. Box 35, Protvino, (Moscow Region), Russian Federation
    \label{SERPUKHOV}}
\titlefoot{J. Stefan Institute, Jamova 39, SI-1000 Ljubljana, Slovenia
    \label{SLOVENIJA1}}
\titlefoot{Laboratory for Astroparticle Physics,
     University of Nova Gorica, Kostanjeviska 16a, SI-5000 Nova Gorica, Slovenia
    \label{SLOVENIJA2}}
\titlefoot{Department of Physics, University of Ljubljana,
     SI-1000 Ljubljana, Slovenia
    \label{SLOVENIJA3}}
\titlefoot{Fysikum, Stockholm University,
     Box 6730, SE-113 85 Stockholm, Sweden
    \label{STOCKHOLM}}
\titlefoot{Dipartimento di Fisica Sperimentale, Universit\`a di
     Torino and INFN, Via P. Giuria 1, IT-10125 Turin, Italy
    \label{TORINO}}
\titlefoot{INFN,Sezione di Torino and Dipartimento di Fisica Teorica,
     Universit\`a di Torino, Via Giuria 1,
     IT-10125 Turin, Italy
    \label{TORINOTH}}
\titlefoot{Dipartimento di Fisica, Universit\`a di Trieste and
     INFN, Via A. Valerio 2, IT-34127 Trieste, Italy
    \label{TRIESTE}}
\titlefoot{Istituto di Fisica, Universit\`a di Udine and INFN,
     IT-33100 Udine, Italy
    \label{UDINE}}
\titlefoot{Univ. Federal do Rio de Janeiro, C.P. 68528
     Cidade Univ., Ilha do Fund\~ao
     BR-21945-970 Rio de Janeiro, Brazil
    \label{UFRJ}}
\titlefoot{Department of Radiation Sciences, University of
     Uppsala, P.O. Box 535, SE-751 21 Uppsala, Sweden
    \label{UPPSALA}}
\titlefoot{IFIC, Valencia-CSIC, and D.F.A.M.N., U. de Valencia,
     Avda. Dr. Moliner 50, ES-46100 Burjassot (Valencia), Spain
    \label{VALENCIA}}
\titlefoot{Institut f\"ur Hochenergiephysik, \"Osterr. Akad.
     d. Wissensch., Nikolsdorfergasse 18, AT-1050 Vienna, Austria
    \label{VIENNA}}
\titlefoot{Inst. Nuclear Studies and University of Warsaw, Ul.
     Hoza 69, PL-00681 Warsaw, Poland
    \label{WARSZAWA}}
\titlefoot{Now at University of Warwick, Coventry CV4 7AL, UK
    \label{WARWICK}}
\titlefoot{Fachbereich Physik, University of Wuppertal, Postfach
     100 127, DE-42097 Wuppertal, Germany \\
\noindent
{$^\dagger$~deceased}
    \label{WUPPERTAL}}
\addtolength{\textheight}{-10mm}
\addtolength{\footskip}{5mm}
\clearpage
\headsep 30.0pt
\end{titlepage}
%
\pagenumbering{arabic} 
\setcounter{footnote}{0} %
\large

%
%









\section{Introduction}
\label{sec:introduction}

\begin{figure}
\begin{center}
\begin{tabular}{ccc} 
\epsfig{file=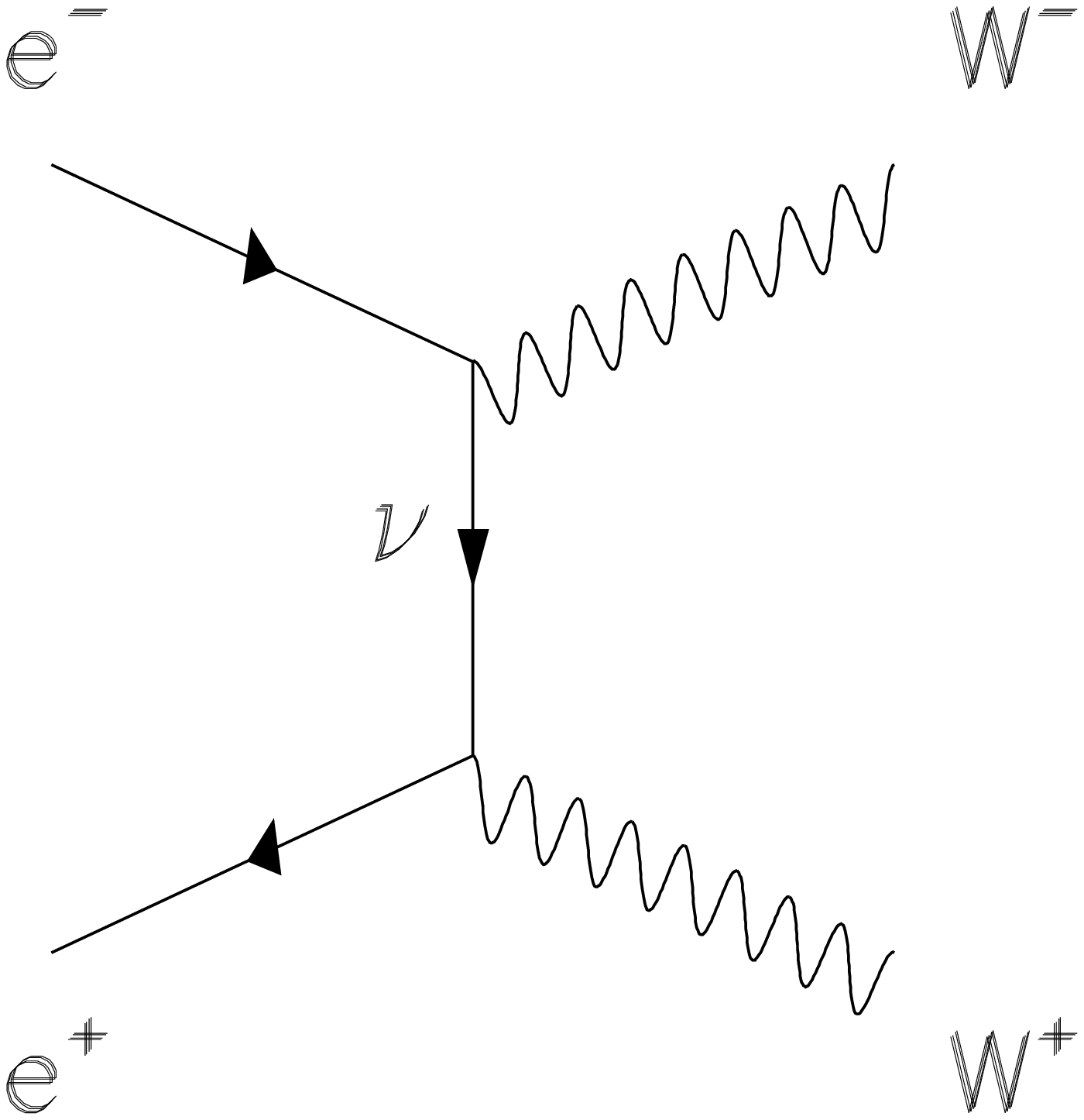,width=4cm} & \epsfig{file=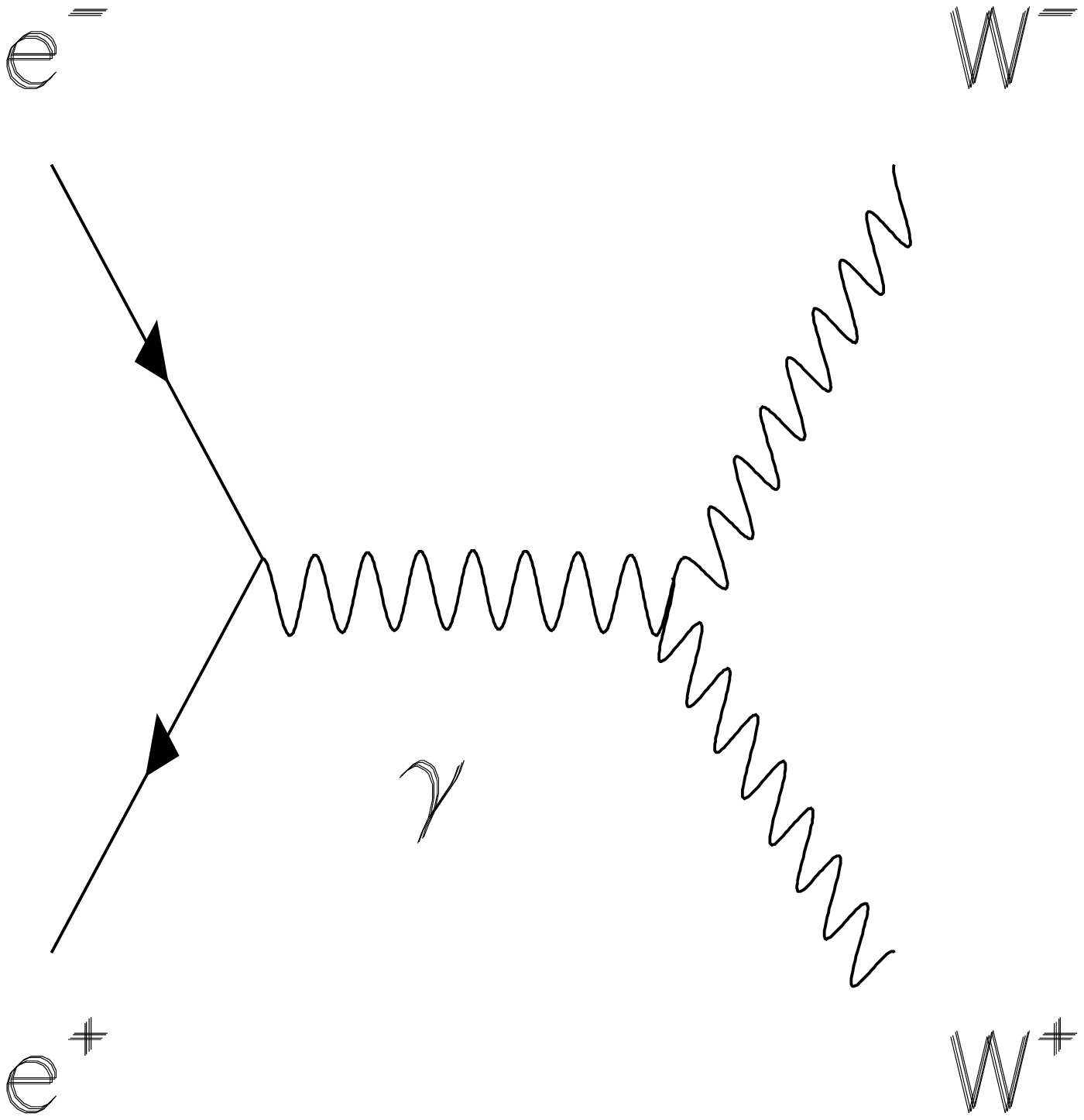,width=4cm} & \epsfig{file=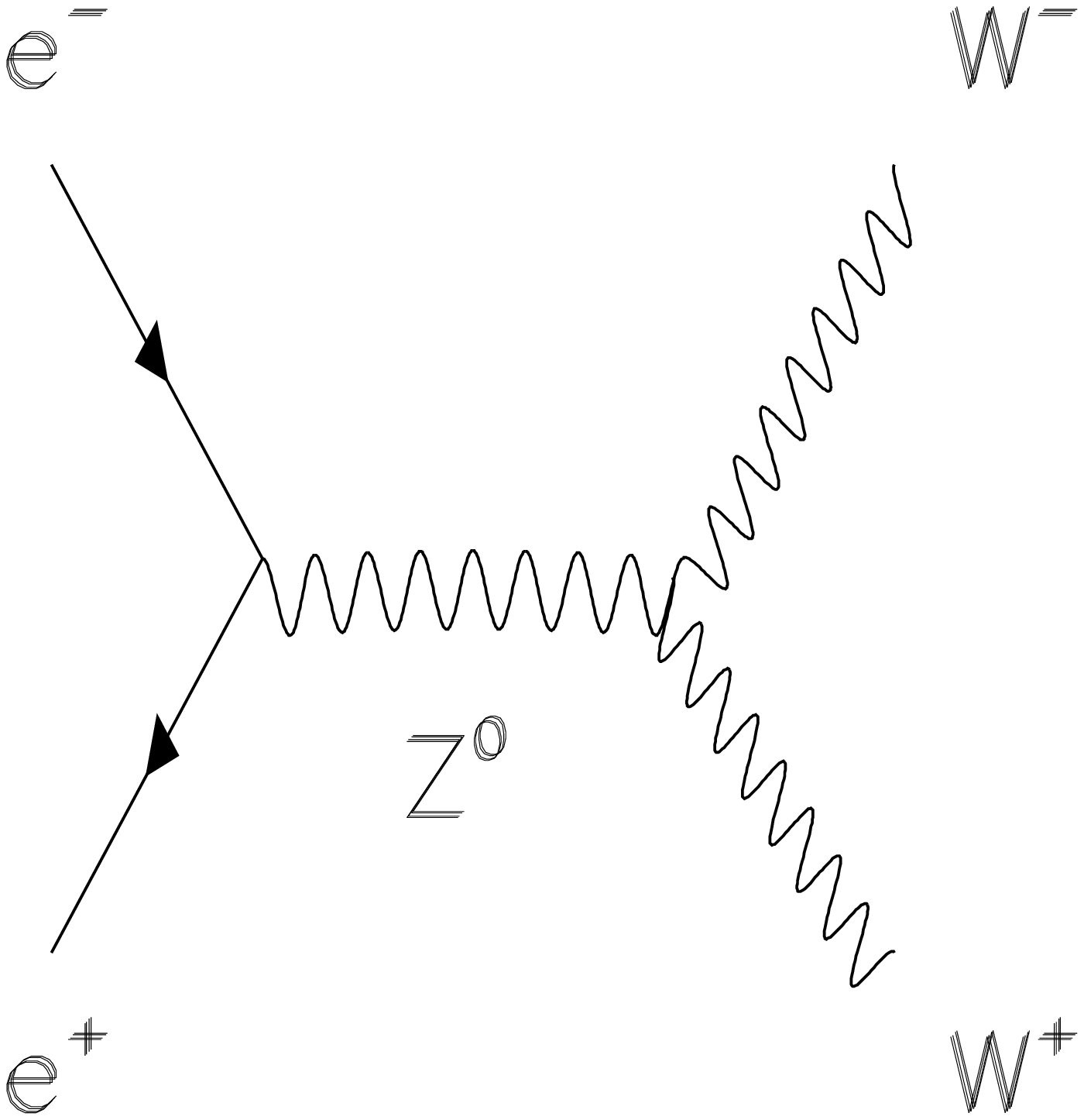,width=4cm} \\
\end{tabular}
\caption[]{CC03 diagrams
} 
\label{fig:diagrams}
\end{center}
\end{figure}

This paper reports on a study of W boson polarisations and measurements of 
Triple Gauge Couplings (TGC's) in the reaction $e^+ e^- \rightarrow W^+ W^-$, using data taken by the DELPHI experiment at LEP at centre-of-mass energies between 189 and 209 GeV.
The amplitude of the reaction $e^+ e^- \rightarrow W^+ W^-$ results from t-channel neutrino and s-channel $\gamma$ and Z exchange and is dominated by the lowest order, so-called CC03, diagrams (see figure \ref{fig:diagrams}).
The s-channel diagrams contain trilinear $\gamma W^+ W^-$ and $Z W^+ W^-$ gauge boson couplings whose possible deviations from the predictions of the Standard Model (anomalous TGC's), due to the effects of new physics, have been extensively discussed in the literature and are for instance described in references \cite{ref:yellow-report-TGC,ref:gaemers-hagiwara,ref:bilenky,ref:gounaris}. 
The decay angles of the 
charged lepton in the $\rm W^-(W^+)$ rest
frame are used to extract the single W CC03 Spin Density Matrix (SDM) elements as a function of the $\rm W^-$ production angle with respect to the $\rm e^-$ beam direction. 
The method of projection operators described in reference  \cite{ref:gounaris} is used.
Measurements of the SDM elements in the reaction $ e^+ e^- \rightarrow W^+ W^-$ have been reported by OPAL \cite{ref:OPAL-SDM}.

The diagonal SDM elements have been 
used to obtain the  differential cross-sections for longitudinally  polarised 
W bosons. 
The study of the longitudinal cross-section  is particularly interesting as this degree of freedom of the W only arises in the Standard Model through the electroweak symmetry breaking mechanism. 
Measurements of the W polarisations at LEP have been reported previously by 
OPAL \cite{ref:OPAL-SDM} and L3 \cite{ref:L3-pola}.
The imaginary parts of the off-diagonal $\rm W^+$ and $\rm W^-$ SDM  elements should vanish in the Standard Model and are particularly sensitive to  CP-violation~\cite{ref:gounaris2}.
  Previous studies of CP-violation in the reaction $ e^+ e^- \rightarrow W^+ W^-$ have been performed by 
ALEPH \cite{ref:ALEPH-tgc}, DELPHI \cite{ref:DELPHI-cptgc} and OPAL \cite{ref:OPAL-SDM}.

Fits were performed to SDM elements measured as a function of the $\rm W^-$ production angle with respect to the $\rm e^-$ beam direction in order to extract CP-conserving and CP-violating charged Triple Gauge boson Couplings.
In this paper 
the theoretical framework described in \cite{ref:yellow-report-TGC}, based on the references given in \cite{ref:gaemers-hagiwara}, is used.
The effective Lagrangian  containing only the lowest dimension operators (up to dimension six;
terms of higher dimensions should be negligible at LEP energies \cite{ref:yellow-report-TGC})
and describing the most general Lorentz invariant $WWV$ vertex, with $V = \gamma$ or $Z$, contains 14 terms with 14 corresponding couplings, $g_1^V, \kappa_V, \lambda_V, g_4^V, g_5^V, \tilde{\kappa}_V, \tilde{\lambda}_V$, representing the annihilation through  the two virtual bosons ( $\gamma$ and $ Z$). 
Assuming $SU(2)_L \times U(1)_Y$ gauge invariance to be preserved, the following constraints between coupling constants are 
obtained \cite{ref:yellow-report-TGC,ref:bilenky}:
\begin{eqnarray}
\Delta \kappa_Z & = & \Delta g_1^Z - \tan^2 \theta_W \cdot \Delta \kappa_{\gamma} \\
\lambda_Z  & = & \lambda_{\gamma}\\
\tilde{\kappa}_Z & = & - \tan^2 \theta_W \cdot  \tilde{\kappa}_{\gamma} \\
\tilde{\lambda}_Z &  = &  \tilde{\lambda}_{\gamma}
\end{eqnarray}

\noindent
with $\Delta \kappa_Z = \kappa_Z - 1$, $\Delta \kappa_{\gamma} = \kappa_{\gamma} - 1$, $\Delta g_1^Z = g_1^Z - 1 $ and $\theta_W$ the weak mixing angle.

Electromagnetic gauge invariance implies  that $g_1^{\gamma} = 1$ and $g_5^{\gamma} = 0 $ for on-shell photons ($q^2=0$) \cite{ref:yellow-report-TGC} . 
In the following the possible $q^2$-dependence of all the TGC's will be assumed to be negligible and we 
set\footnote
{The parameters $g_1^{\gamma}, \kappa_{\gamma}$ and $\lambda_{\gamma}$ are related to the charge $Q_W$, the magnetic dipole moment $\mu_W$ and the electric quadrupole moment $q_W$ of the $W^+$ with: 
\\
$Q_W = e g_1^{\gamma}$, $\mu_W = \frac{e}{2 m_W} (g_1^{\gamma} + \kappa_{\gamma} + \lambda_{\gamma}) $ and $q_W = - \frac{e}{m_W^2} (\kappa_{\gamma} - \lambda_{\gamma})$. }
$g_1^{\gamma} = 1$ and assume that the CP-violating coupling $g_4^{\gamma} = 0$ and that $g_5^{\gamma} = g_5^Z = 0$ at all $q^2$. 
These last two coupling constants, although CP-conserving, correspond to the only terms violating both C- and P-symmetry in the  Lagrangian considered in this analysis.

With these assumptions, the number of independent coupling parameters can be reduced to six, three of which correspond to CP-conserving interactions ($\Delta g_1^Z, \Delta \kappa_{\gamma}$ and $\lambda_{\gamma}$), the remaining three being CP-violating ($g_4^Z, \tilde{\kappa}_Z$ and $\tilde{\lambda}_Z$). 
 In the Standard Model (SM) all these parameters are expected to be zero at tree level. 
Hence $\Delta g_1^Z$ and $ \Delta \kappa_{\gamma}$ explicitly parameterise possible anomalous deviations of the couplings $g_1^Z$ and $\kappa_{\gamma}$ from their Standard Model values.

Triple Gauge Couplings have been measured by the four LEP experiments, ALEPH \cite{ref:ALEPH-tgc}, DELPHI \cite{ref:DELPHI-tgc}, L3 \cite{ref:L3-tgc} and OPAL \cite{ref:OPAL-tgc}. 
The most recent results 
from DELPHI on CP-conserving TGC's \cite{ref:DELPHI-tgc} were derived from data taken at centre-of-mass energies ranging from 189 to 209 GeV. 
Hadronic as well as leptonic decay channels of the 
W bosons 
were considered using methods based on angular observables characterising both W production and decay. 
Measurements of CP-violating TGC's analogous to those described in this paper have been made by OPAL 
\cite{ref:OPAL-cptgc}, while results from a different fit method have been published by ALEPH \cite{ref:ALEPH-tgc}.

The selection of semi-leptonic 
$ e^+ e^- \rightarrow W^+ W^- \rightarrow l \nu q \bar{q} (l = e,  \mu)$ 
events and the corrections for efficiency, resolution and purity are given in section \ref{sec:datasample}. 
Section \ref{sec:sdm} discusses the determination of the single W SDM elements, the estimation of the fraction of longitudinally polarised Ws and the study of CP-violating effects on the imaginary elements. 
Section \ref{sec:systematics} is devoted to the estimation of the systematic errors on the SDM's. 
The TGC fits are described in section \ref{sec:tgc}. 
A global summary is given in section \ref{sec:summary}.

\section{Data sample and Monte Carlo simulation}
\label{sec:datasample}

For this analysis the data taken by DELPHI at centre-of-mass energies between
189 and 209 GeV were used. 
The DELPHI detector and its performance are described in reference \cite{delphi-det}.
The data consist of events of the type 
$ e^+ e^- \rightarrow W^+ W^- \rightarrow l \nu q \bar{q} (l = e,  \mu)$.
In order to take the energy dependence of the measurements into account, the data 
 were grouped into three samples: 154 $\rm pb^{-1}$ taken in 1998 at  189 GeV, 218 $\rm pb^{-1}$ taken in 1999 at energies between 192 and 202 GeV (mean 198 GeV) and 149 $\rm pb^{-1}$ taken in 2000 at energies in the range 204 to 209 GeV (mean 206 GeV).

Events were selected in which one W decayed into a $ e \nu$ or $\mu \nu$ pair while the other W 
decayed into a pair of quarks. 
These events are characterised by one isolated electron or muon, two hadronic jets and missing momentum coming from the neutrino. 
The major background comes from $q\bar{q}\tau\nu$ final states, from $q \bar{q} (\gamma)$ production and from neutral current four-fermion final states containing two quarks and two leptons.

After a loose preselection, an Iterative Discriminant Analysis (IDA)
was used to make the final selection. 
This part of the event selection is identical to the procedure used to measure the 
WW production cross-sections \cite{DELPHI-xsec}. 
Events were selected with a cut on the output of the IDA, chosen to optimise
the product of efficiency and purity for each channel. 
Events were first passed to the $q\bar{q}\mu\nu$ selection; if they were not selected,
they were passed to the $q\bar{q}e\nu$ selection; if they were still not selected, 
they were then finally passed to the $q\bar{q}\tau\nu$ selection 
for possible inclusion or rejection. 
In this analysis
only the events tagged as $q\bar{q}e\nu$ or $q\bar{q}\mu\nu$ were retained.

A three-constraint kinematic fit was then applied in which  the masses of the two W candidates were constrained to be equal to a reference mass (80.35 $\rm GeV/c^2$). 
A cut was applied on the $\rm \chi^2$ probability of this fit at 0.005. 
Events for which the angle between the lepton track and the beam axis was less than 20$^{\circ}$ 
 were rejected to remove leptons with  poor charge measurement. 

The integrated luminosity used is 520 $\rm pb^{-1}$, corresponding to data taking runs in which the subdetectors which were essential for this analysis were fully operational. A total of 1880 $ l \nu q \bar{q}$ events was selected. The data were analysed separately for each of the three years. A breakdown of the collected statistics for different energies, as well as the mean energy for each sample, are given in table \ref{tab:statistics}, with other details.

The signal refers to the WW-like CC03 diagrams leading to 
$l \nu q \bar{q}$ final states \cite{ref:gounaris}. 
The efficiencies and purities were estimated by Monte Carlo methods with the 
WPHACT \cite{ref:wphact} program (charged and neutral current four-fermion events), 
and KK2F \cite{ref:KK2F} ($q \bar{q} (\gamma)$ event generator) at energies of 188.6, 199.5 and 206.0 GeV. 
The hadronisation of quarks was simulated with the JETSET \cite{ref:jetset} package.
To account for the full $\cal{O}$($\alpha$) radiative corrections the generated charged current events were reweighted following the procedure 
described in
\cite{ref:dpaweighting}. 
The CC03 selection efficiency  was around 70\%
while the purity was around 92\%. Both were roughly energy independent as shown in table
\ref{tab:statistics}. 


To obtain the SDM elements the selected events were corrected for the  acceptance, the angular resolutions and the sample purity. 
The correction factors were obtained from samples of simulated events 
with sizes 
given in table \ref{tab:statistics}.

\begin{figure}
\begin{center}
\epsfig{file=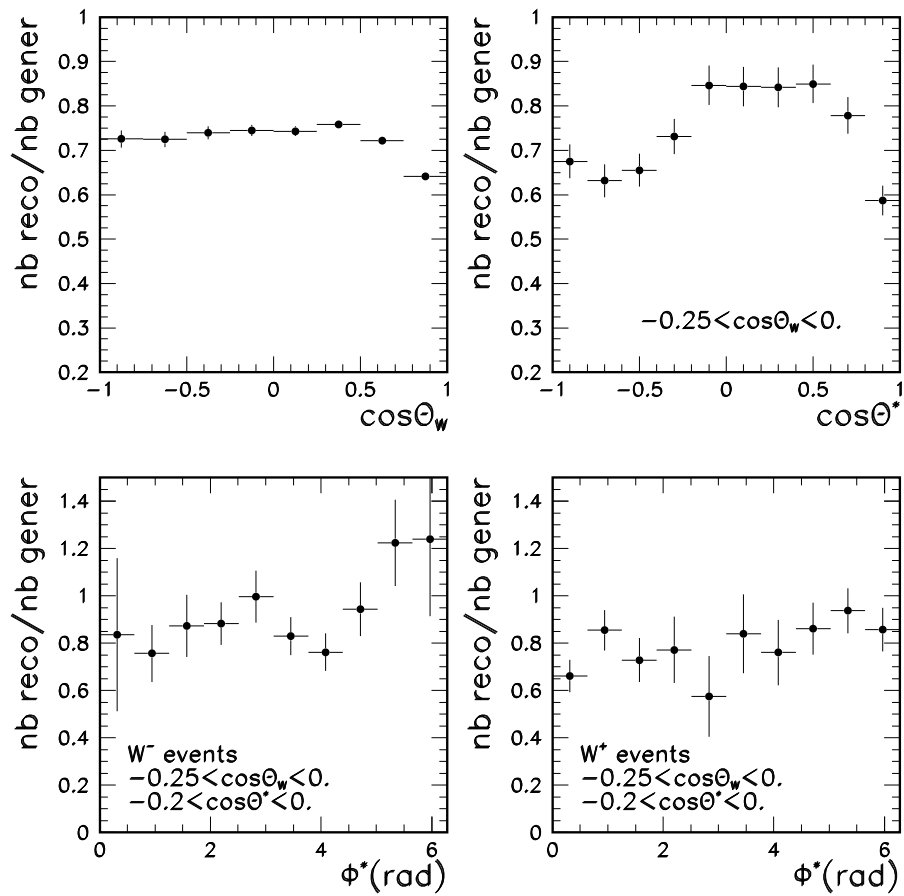 ,width=14cm}
\caption[]{Efficiency as function of $\cos \Theta_W, \cos\theta^*$ and $\phi^*$ at 199.5 GeV, obtained from simulated events.
} 
\label{fig:efficiencies}
\end{center}
\end{figure}

The selection efficiency was calculated as a function of the $\rm W^-$ production angle  $\cos \Theta_W$ and the lepton decay angles $\cos \theta^*$ and $\phi^*$. 
The lepton decay angles are defined in the W rest frame as shown in figure \ref{fig:kine}.
The efficiency is defined as the number of reconstructed events divided by the number of generated events in a given angular interval.
Since the signal refers to the CC03 diagrams only, each event was reweighted by the ratio of the square of the matrix element for the CC03 diagrams only to the square of the matrix element for the full set of diagrams leading to $qqe\nu$ and $qq\mu\nu$ final states, including the full $\cal{O}$($\alpha$) radiative corrections.
 The events were divided in 8 equal bins of $\cos \Theta_W$, in 10 equal bins of $\cos \theta^*$ and in 10 equal bins of $\phi^*$. The corrections were computed in each of these three-dimensional bins. The average number of generated events in a bin was 80 and about 7\% of the bins were populated by less than 10 events.
Examples of the efficiency distributions at 199.5 GeV are shown in figure \ref{fig:efficiencies}. 

The typical resolution on the measured $\cos\Theta_W$, after the 3C 
kinematic fit, was found to be 0.04, much smaller than the bin size of 0.25. 
For about 17\% of the events the reconstructed $\cos\Theta_W$ deviates from the generated value by more than 0.125. 
Because of the definition of the selection efficiency as the convolution of efficiency and migration,  correlations between neighbouring $\cos\Theta_W$ bins are expected after the correction procedure. 
A study of simulated events shows that between 70\% and 90\% of the events are reconstructed in the correct bin, and that the remaining events are nearly all reconstructed in the directly neighbouring intervals.
The typical resolution on the measured $\rm cos\theta^*$ was  0.05, while it 
was 0.08 radians on the measured $\rm \phi^*$. 
This has to be compared to the bin widths of 0.2 and 0.628 radians respectively. 

\begin{figure}
\begin{center}
\epsfig{file=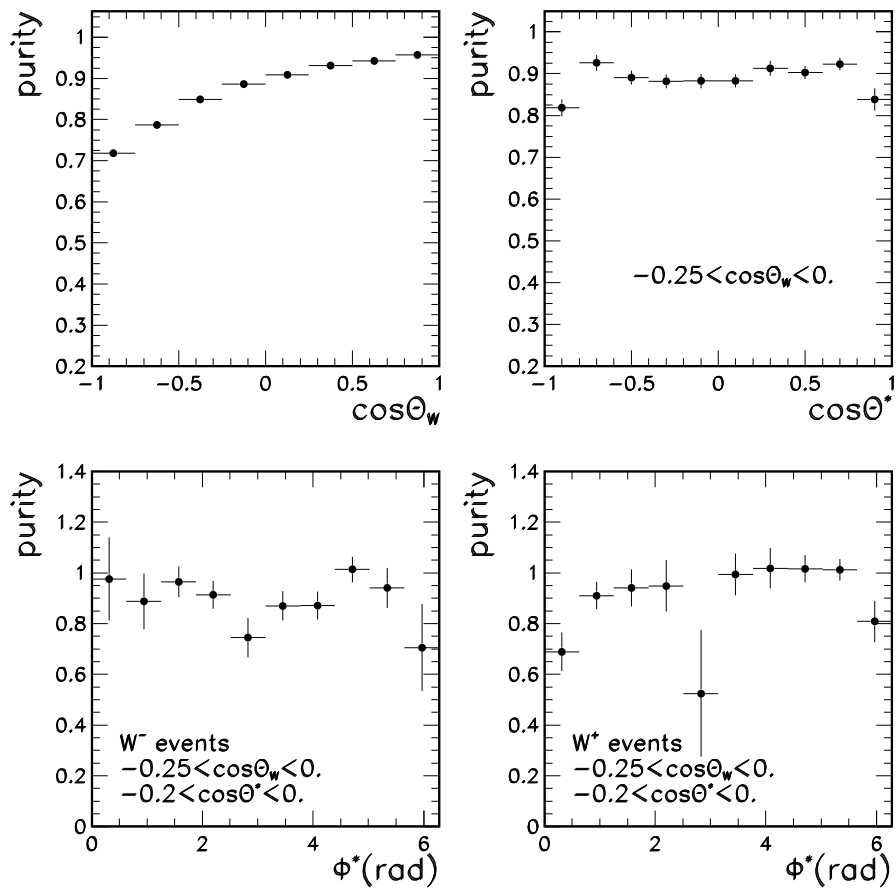 ,width=14cm}
\caption[]{Purity as function of $\cos \Theta_W, \cos\theta^*$ and $\phi^*$ at 199.5 GeV, obtained from simulated events.
} 
\label{fig:purities}
\end{center}
\end{figure}

The purity with respect to CC03 $e/\mu$ production was calculated as a function of the three relevant angles with the same binning as used for the efficiencies.
 To estimate the signal contribution, the WW events were reweighted to obtain `CC03 events' as explained above for the efficiency estimation. 
To estimate the background from $\tau \nu q \bar{q}$ and fully hadronic WW final states the events were reweighted to account for full O($\alpha$) radiative corrections.
The small contribution of non-CC03 semi-leptonic e/$\mu$ events was also accounted as background. 
The other background contributions come from $q \bar{q} (\gamma)$ and neutral current four-fermion final states.
Examples of the purity distributions at 199.5~GeV are shown in figure \ref{fig:purities}. 
Effective purities can become slightly greater than 1 due to interference effects between CC03 and higher order diagrams affecting the CC03 reweighting procedure \cite{ref:dpaweighting}

\begin{figure}
\begin{center}
\epsfig{file=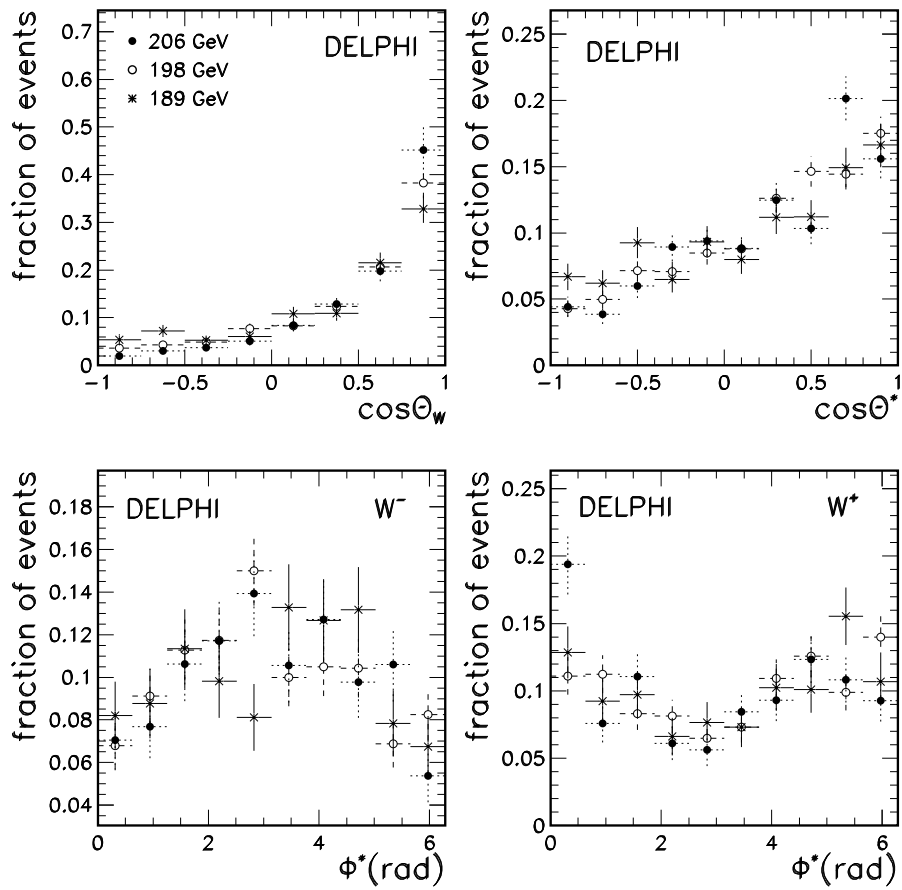 ,width=14cm}
\caption[]{Angular distributions, normalised to one and fully corrected, for data taken at 189 GeV, 198 GeV and 206 GeV.
} 
\label{fig:angular-distributions}
\end{center}
\end{figure}

The fully corrected production and decay angle distributions obtained from the data are shown in figure \ref{fig:angular-distributions} for the three data-taking years. 
The $\cos\Theta_W$ and $\cos\theta^*$ distributions for $\rm W^-$ and $\rm W^+$ events,
with the W  decaying respectively in a negative or positive lepton, have been added together.

\section{Single W Spin Density Matrix and W polarisation}
\label{sec:sdm}

For events of the type
\begin{eqnarray*}
e^+ (\lambda') \ e^-(\lambda) \rightarrow W^+(\tau_+) \ W^-(\tau_-)
\end{eqnarray*}
where $\rm\lambda=\pm \frac{1}{2} \ (\lambda'=-\lambda)$ is the helicity of the electron (positron), $\rm\tau_-=\pm 1,0$ and $\rm\tau_+=\pm 1,0$ are the helicities of the $\rm W^-$ and $\rm W^+$, respectively, the two-body 
spin density matrix (SDM) is defined as \cite{ref:yellow-report-TGC,ref:bilenky,ref:gounaris}:
\begin{equation}
\rho_{\tau_- \tau_-' \tau_+ \tau_+'}(s,\cos\Theta_W) = 
\frac{\sum_{\lambda} F_{\tau_- \tau_+}^{(\lambda)}
F_{\tau_-' \tau_+'}^{*(\lambda)}}{\sum_{\lambda \tau_- \tau_+}|F_{\tau_- \tau_+}^{(\lambda)}|^2}
\end{equation}
with $\cos\Theta_W$ the production angle of the $\rm W^-$ with respect to the
$\rm e^-$ beam and $\rm F_{\tau_- \tau_+}^{(\lambda)}$ the amplitude for the
production of a $\rm W^-$ with helicity $\rm\tau_-$ and a  
$\rm W^+$ with helicity $\rm\tau_+$. 
If only $\rm W^-$ decays are observed we have
\begin{eqnarray*}
\rho^{W^-}_{\tau_- \tau_-'} (s,\cos\Theta_W) = 
      \sum_{\tau_+} \rho_{\tau_- \tau_-' \tau_+ \tau_+}(s,\cos\Theta_W),
   \hspace{1.5cm}   \sum_{\tau_-} \rho^{W^-}_{\tau_- \tau_-} = 1.  
\end{eqnarray*}
In an analogous way, one has:
\begin{eqnarray*}
\rho^{W^+}_{\tau_+ \tau_+'} (s,\cos\Theta_W) = 
      \sum_{\tau_-} \rho_{\tau_- \tau_- \tau_+ \tau_+'}(s,\cos\Theta_W),
\hspace{1.5cm} \sum_{\tau_+} \rho^{W^+}_{\tau_+ \tau_+} = 1.
\end{eqnarray*}

\begin{figure}
\begin{center}
\epsfig{file=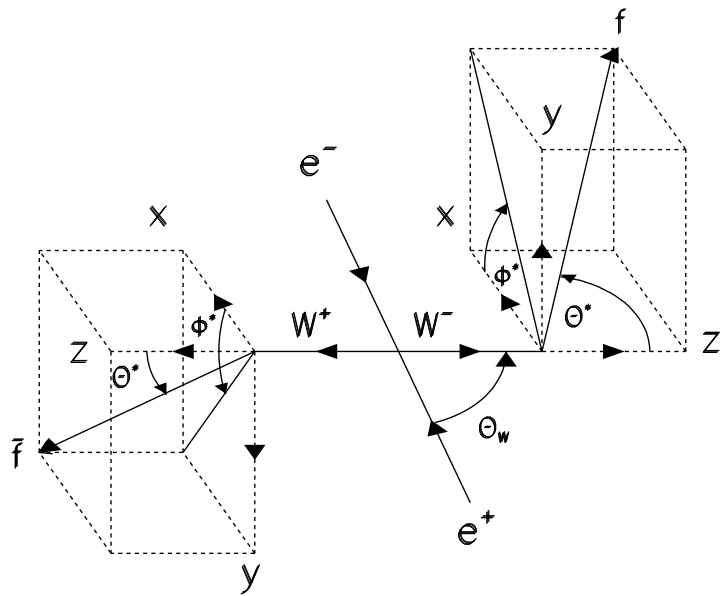 ,width=13cm}
\caption[]{Definition of the $\rm W^-$ production angle $\Theta_W$ and the lepton decay angles $\theta^*$ and $\phi^*$ in the rest frame of the W. 
} 
\label{fig:kine}
\end{center}
\end{figure}

The differential cross-section  for $\rm W^+ W^-$ production with subsequent leptonic
decay of the $\rm W^-$ can be written as:
\begin{eqnarray*}
\frac{d^3 \sigma (e^+ e^- \rightarrow W^+ W^- \rightarrow W^+  \ell^- \bar{\nu})}
{d \cos \Theta_W d\cos \theta^* d \phi^*} \times \frac{1}{BR} =  \hspace{7cm}\\
 \hspace{4cm}  \frac{d \sigma (e^+ e^- \rightarrow W^+ W^-)}{d \cos \Theta_W} 
(\frac{3}{8 \pi})
\sum_{\tau_- \tau_-'} \rho^{W^-}_{\tau_- \tau_-'} (s,\cos\Theta_W)
D_{\tau_- \tau_-'}(\theta^*,\phi^*),
\end{eqnarray*}
where the 
$ D_{\tau_- \tau_-'}(\theta^*,\phi^*)$ 
functions describe the standard (V-A) decay of the $\rm W^-$, 
$\rm (\theta^*,\phi^*)$ 
are the angles of the lepton in the $\rm W^-$ rest frame 
(see figure \ref{fig:kine}) and BR is the 
$ W^- \rightarrow \ell^- \bar{\nu}$ branching fraction. 
The coordinate system in which these angles are
defined is 
that of ref. \cite{ref:bilenky} and corresponds to the one shown in figure \ref{fig:kine}.
This representation of the
differential cross-section  in terms of the spin density matrix is independent of
the specific form of the helicity amplitudes, i.e. of the specific form of the
$\rm W^+ W^-$ production process. The empirical determination of the SDM
elements thus amounts to a model-independent analysis of this process.

A set of projection operators $ \Lambda^{W^-}_{\tau_- \tau_-'}$ can be found
\cite{ref:gounaris} which isolate the corresponding 
$\rm \rho^{W^-}_{\tau_- \tau_-'}$ contributions when integrated over the full
lepton spectrum:
\begin{eqnarray*}
\rho^{W^-}_{\tau_- \tau_-'} = 
\frac{1}{BR \times \frac{d \sigma(e^+ e^- \rightarrow W^+ W^-)}{d \cos \Theta_W}}
\int \frac{d^3 \sigma(e^+ e^- \rightarrow W^+  \ell^- \bar{\nu})}
{d \cos \Theta_W d\cos \theta^* d \phi^*} 
\Lambda^{W^-}_{\tau_- \tau_-'}(\theta^*,\phi^*) d\cos \theta^* d \phi^*.
\end{eqnarray*}
The SDM elements for $\rm W^+$ production are obtained in a similar way.

For a CP-invariant interaction, such as in the standard 
$\rm SU(2)_L \times U(1)_Y$
theory, the SDM elements of the produced $\rm W^+$ and $\rm W^-$ are related via
\cite{ref:gounaris2}:
\begin{equation}
\rho^{W^-}_{\tau_- \tau_-'} (s,\cos \Theta_W) = 
\rho^{W^+}_{-\tau_- -\tau_-'} (s,\cos \Theta_W).
\label{equ:rho-wmwp}
\end{equation}

\noindent
The magnitude of any difference between the left-hand and right-hand sides of 
(\ref{equ:rho-wmwp}) constitutes a direct measure of the strength of a possible
CP-violating interaction. 
At tree level, invariance under CPT transformations also implies the validity of
relations (\ref{equ:rho-wmwp}) when applied to the real parts of the SDM, while
for the imaginary parts, CPT invariance leads to the relation:
\begin{eqnarray}
Im \rho^{W-}_{\tau_- \tau_-'} + Im \rho^{W+}_{-\tau_- -\tau_-'} = 0.
\label{equ:rho-wmp}
\end{eqnarray}
\noindent
Thus a violation of CP-invariance in WW production can best be investigated by looking for inequality of the imaginary parts of the SDM in (\ref{equ:rho-wmwp}), i.e. by testing the relations:
\begin{eqnarray}
\label{equ:difimm1}
Im \rho^{W^-}_{\tau_- \tau_-'} - Im \rho^{W^+}_{-\tau_- -\tau_-'} = 0.
\label{equ:difimm3}
\end{eqnarray}
Relations (\ref{equ:rho-wmp}) and (\ref{equ:difimm1}) result in the fact that the imaginary parts of the SDM should vanish.

Experimentally the SDM elements were obtained from the relation
\begin{equation}
\rho_{\tau_- \tau_-'}^{W^{\pm}}(s,\cos\Theta_W) = 
\frac{1}{\sum_{j=1}^{N_i}w_j} \sum_{j=1}^{N_i} 
w_j \Lambda_{\tau_- \tau_-'}^{W^{\pm}}(\cos\theta^*_j,\phi^*_j),
\label{equ:rho-exp}
\end{equation}
where $\rm N_i$ is the number of selected events in a given $ \cos\Theta_W$ bin.
Each event was weighted
with a correction factor $w_j$ dependent on $( \cos\Theta_W,\cos\theta^*,\phi^*)$ as explained in section \ref{sec:datasample}, to account for detector acceptance,
bin migration 
and sample purity.

\begin{figure}
\begin{center}
\epsfig{file=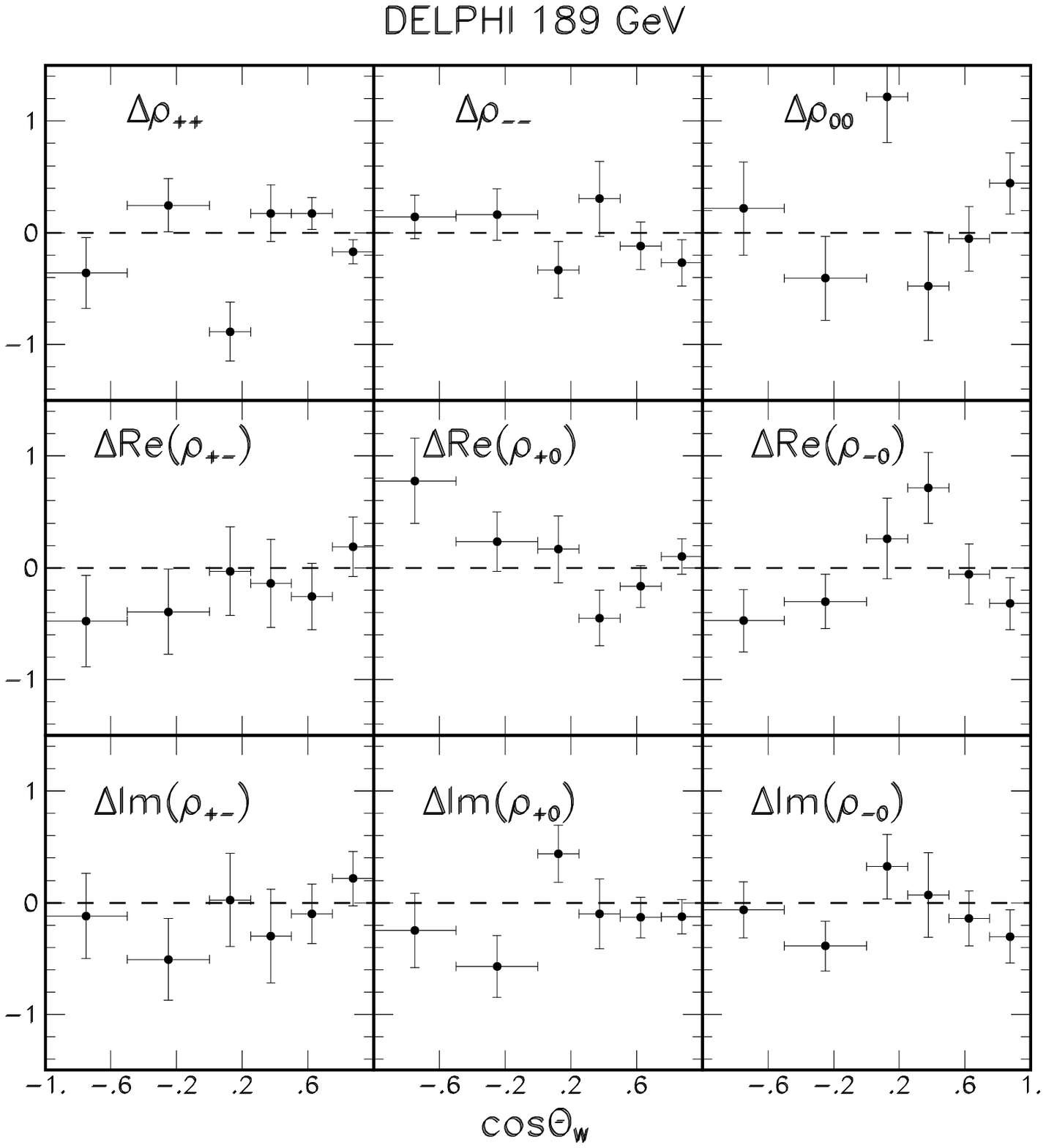 ,width=14cm}
\caption[]{Difference $\Delta \rho_{\tau \tau'} = \rho^{W^-}_{\tau \tau'} (s,\cos \Theta_W) - 
\rho^{W^+}_{-\tau -\tau'} (s,\cos \Theta_W)$ (see equation (\ref{equ:rho-wmwp})), with statistical errors, measured with the data taken at 189 GeV, corrected for detector acceptance and sample purity as explained in the text.  
} 
\label{fig:rho-wmp-98}
\end{center}
\end{figure}

\begin{figure}
\begin{center}
\epsfig{file=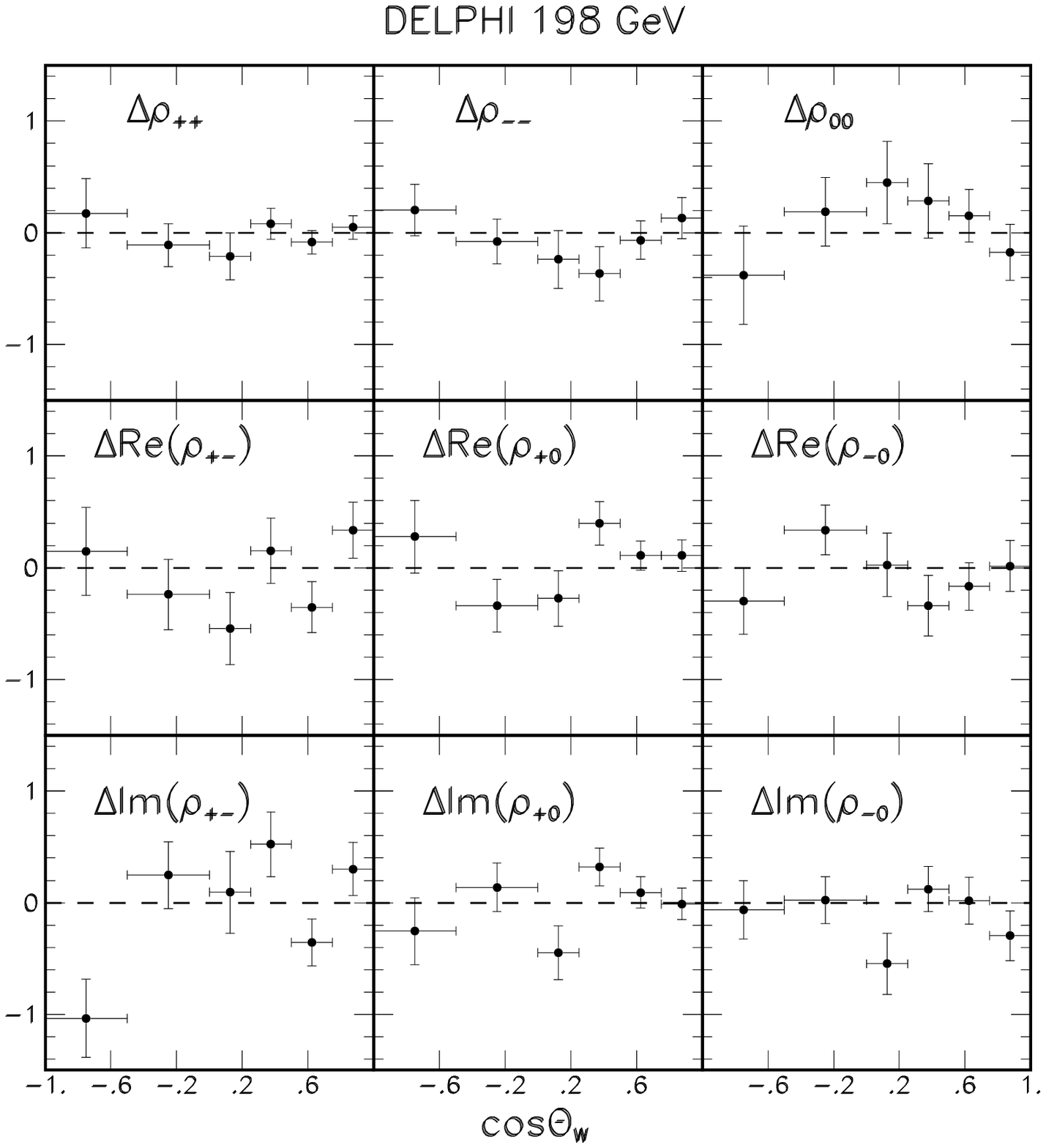 ,width=14cm}
\caption[]{Difference $\Delta \rho_{\tau \tau'} = \rho^{W^-}_{\tau \tau'} (s,\cos \Theta_W) - 
\rho^{W^+}_{-\tau -\tau'} (s,\cos \Theta_W)$ (see equation (\ref{equ:rho-wmwp})), with statistical errors, measured with the data taken at 198 GeV, corrected for detector acceptance and sample purity as explained in the text. 
} 
\label{fig:rho-wmp-99}
\end{center}
\end{figure}

\begin{figure}
\begin{center}
\epsfig{file=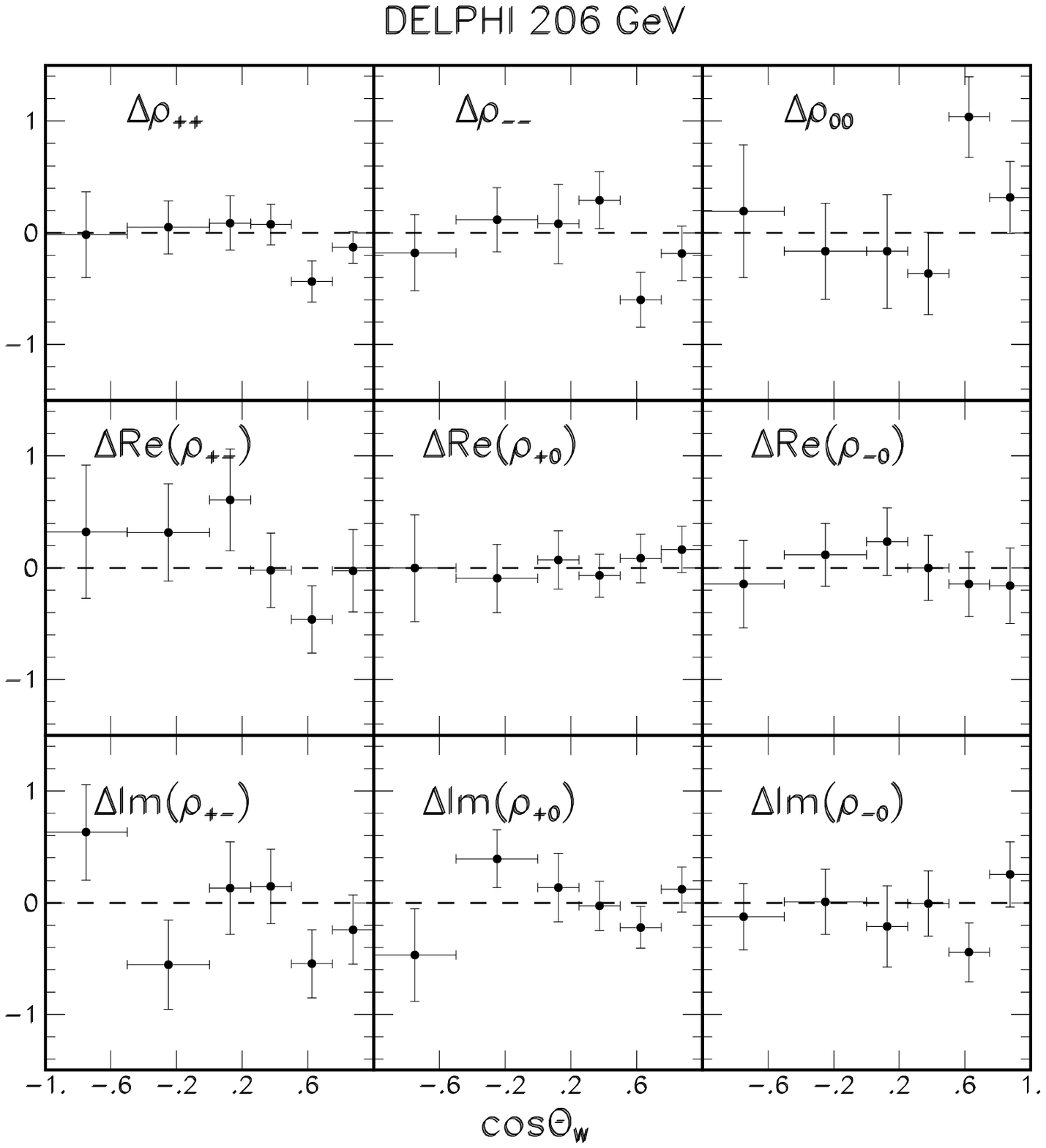 ,width=14cm}
\caption[]{Difference $\Delta \rho_{\tau \tau'} = \rho^{W^-}_{\tau \tau'} (s,\cos \Theta_W) - 
\rho^{W^+}_{-\tau -\tau'} (s,\cos \Theta_W)$ (see equation (\ref{equ:rho-wmwp})), with statistical errors, measured with the data taken at 206 GeV, corrected for detector acceptance and sample purity as explained in the text. 
} 
\label{fig:rho-wmp-00}
\end{center}
\end{figure}

\begin{figure}
\begin{center}
\epsfig{file=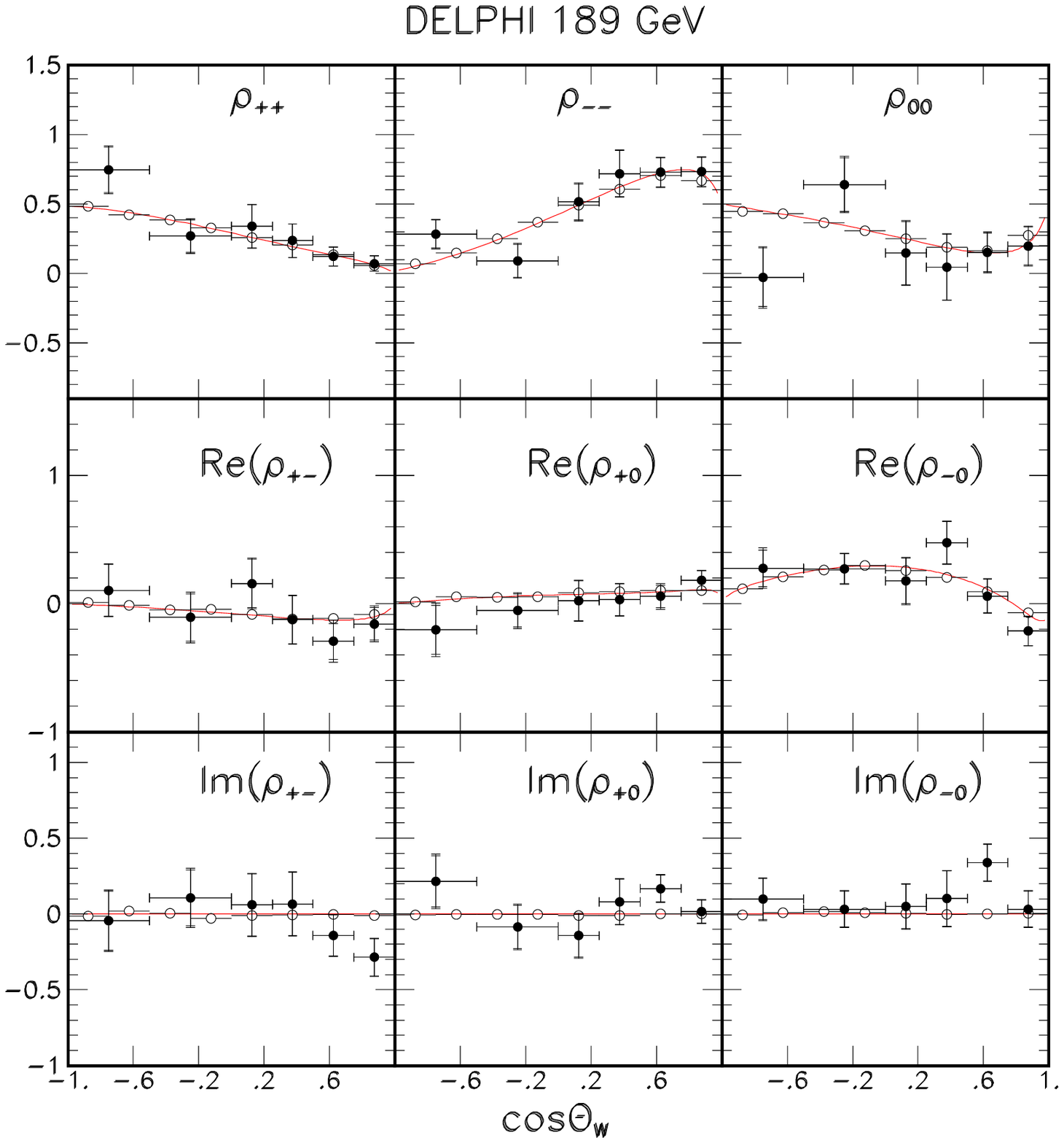 ,width=14cm}
\caption[]{
Averages of $\rm W^+$ and $\rm W^-$
SDM elements, with statistical and total errors, measured with the data taken 
at 189 GeV (black dots), corrected for detector acceptance and sample purity as explained in the text. The full line shows the tree level SM prediction calculated with the analytical expression from ref. \protect{\cite{ref:gounaris}}. The open circles are the SM tree level predictions obtained with the WPHACT MC at generator level. 
} 
\label{fig:rho-data-98}
\end{center}
\end{figure}

\begin{figure}
\begin{center}
\epsfig{file=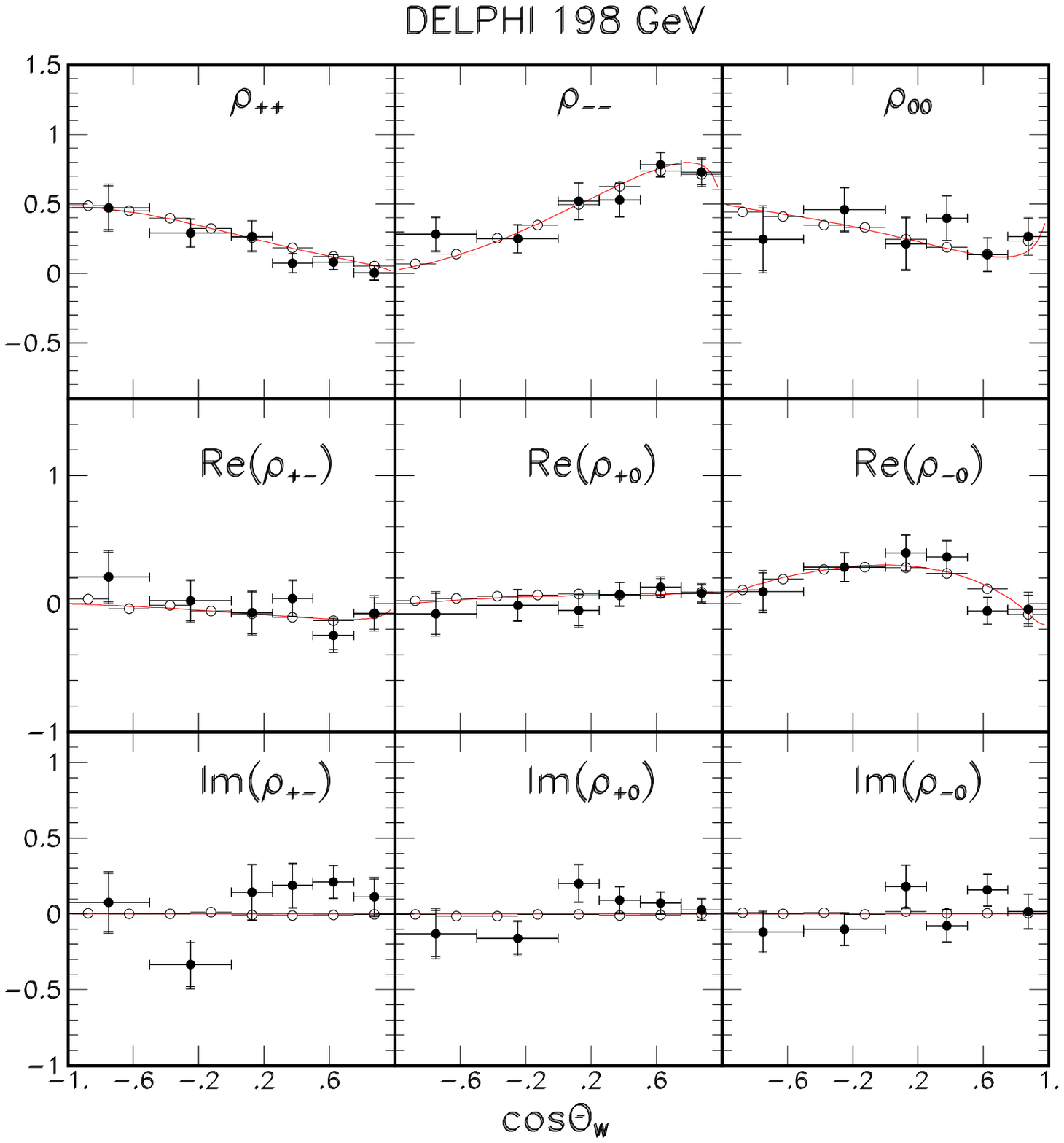 ,width=14cm}
\caption[]{
Averages of $\rm W^+$ and $\rm W^-$
SDM elements, with statistical and total errors,  measured with the data taken 
at an energy of 198 GeV (black dots), corrected for detector acceptance and sample purity as explained in the text. The full line shows the tree level SM prediction calculated with the analytical expression from ref. \protect{\cite{ref:gounaris}}. The open circles are the SM tree level predictions obtained with the WPHACT MC at generator level. 
}
\label{fig:rho-data-99}
\end{center}
\end{figure}

\begin{figure}
\begin{center}
\epsfig{file=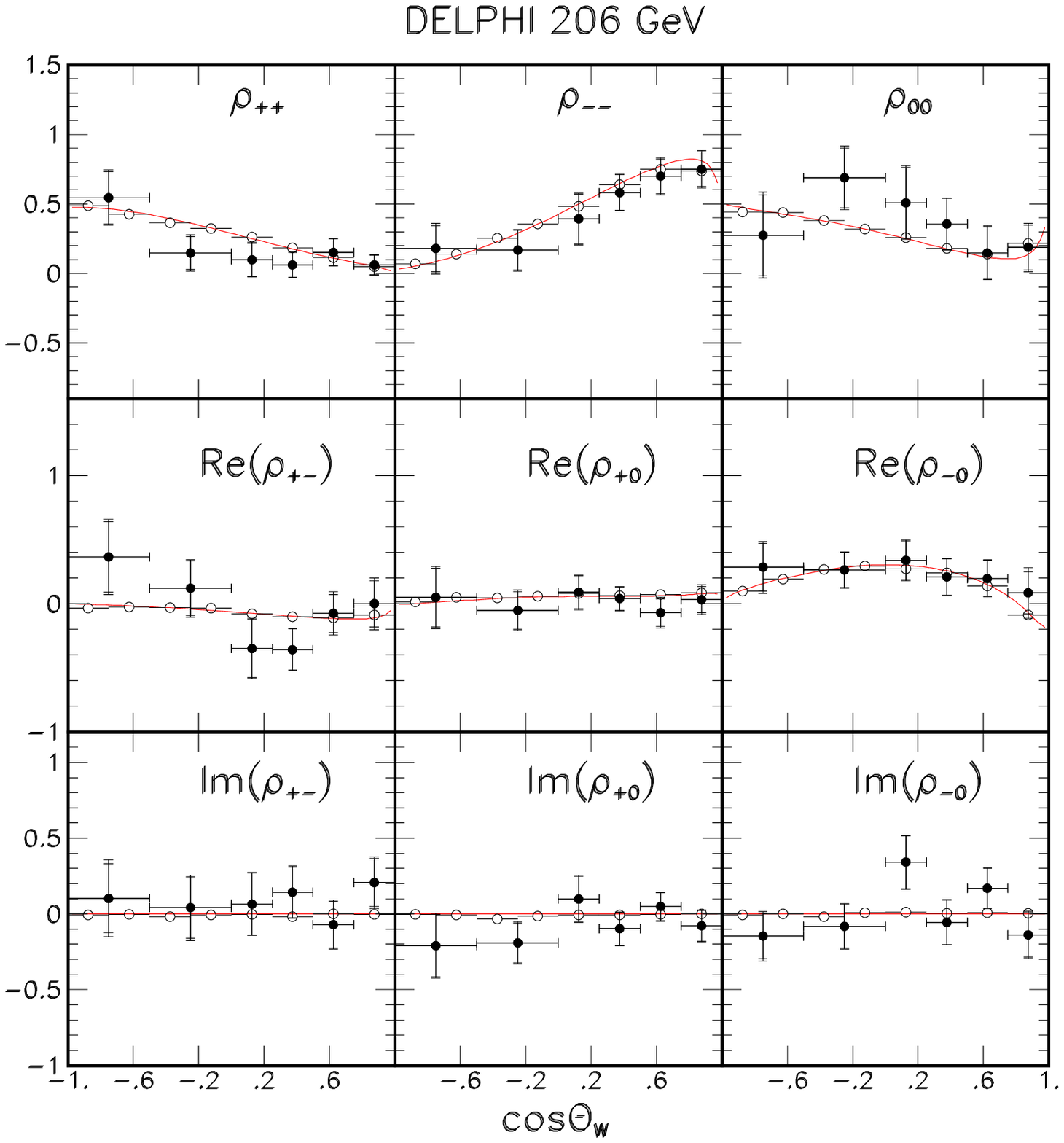 ,width=14cm}
\caption[]{
Averages of $\rm W^+$ and $\rm W^-$
SDM elements, with statistical and total errors,  measured with the data taken at 
 an energy of 206 GeV (black dots), corrected for detector acceptance and sample purity as explained in the text. The full line shows the tree level SM prediction calculated with the analytical expression from ref. \protect{\cite{ref:gounaris}}. The open circles are the SM tree level predictions obtained with the WPHACT MC at generator level.
}
\label{fig:rho-data-00}
\end{center}
\end{figure}

The event sample was divided into 8 equal bins of $ \cos\Theta_W$. 
As the $\rm W^-$ production occurs mainly in the forward direction with respect to the $\ e^-$ beam, and the experimental statistics available are rather restricted, 75\% of the $\cos \Theta_W$ bins 
in the backward region have less than 20 events 
when the $\cos \Theta_W$ values are sampled in eight equal bins.
From WPHACT Monte Carlo studies of a large number (250) of data-sized samples simulated at energies of 189, 200 and 206 GeV, it appears that
the number of events per bin should be at least
about 20 to allow a reliable extraction of Triple Gauge Couplings from the data. 
In  order to reach this goal, the SDM elements were 
redetermined in two equal-sized $\cos \Theta_W$ bins for $\rm W^-$ bosons produced in the backward region.
Figures \ref{fig:rho-wmp-98}, \ref{fig:rho-wmp-99} and \ref{fig:rho-wmp-00} show that the SDM elements computed for $\rm W^+$ and $\rm W^-$ separately are compatible with relation (\ref{equ:rho-wmwp}) imposed by CP-invariance. Only statistical errors are displayed as systematic effects are expected to be small compared to statistical fluctuations (see section \ref{sec:systematics}) and 
are similar for $\rm W^+$ and $\rm W^-$ bosons.
The measurements of the SDM elements are 
shown in figures \ref{fig:rho-data-98}, \ref{fig:rho-data-99}, and \ref{fig:rho-data-00}
for the three data samples taken in 1998, 1999 and 2000 separately.
As the SDM elements computed for $\rm W^+$ and $\rm W^-$ separately are compatible, 
CP-invariance is assumed in these plots and 
both the $\rm W^+$  and $\rm W^-$ leptonic decays were used to compute the $\rm W^-$ SDM elements, based on relation (\ref{equ:rho-wmwp}).
The predictions from  Standard Model
signal events 
(about 50000 $\rm pb^{-1}$ at each energy simulated with WPHACT) are also shown
 together with the results from the analytical calculations used in the TGC fits (see section \ref{sec:tgc}). 
The measured values agree with the SM expectation 
at all energies considered. 
Indeed, the $\chi^2$ values for comparison with the analytical calculation, and taking into account the SDM elements in the 6 bins as shown in the figures \ref{fig:rho-data-98} to \ref{fig:rho-data-00}, are respectively 45.3 (189 GeV), 43.5 (198 GeV) and 35.8 (206 GeV) for 48 degrees of freedom. 
In the calculation of the  $\chi^2$ the 
linear constraints on the diagonal elements were taken into account by removing the element $\rho^{++}$, and the full covariance matrix based on the statistical and systematic errors as explained in section \ref{sec:tgc}, was used. The corresponding $\chi^2$ probabilities are 58.2\%, 65.9\% and 90.2\% respectively. 

In figures \ref{fig:rho-data-98}, \ref{fig:rho-data-99} and \ref{fig:rho-data-00} a comparison is made of the CC03 SDM elements calculated with WPHACT (open dots) and those obtained with the expressions from ref.\cite{ref:gounaris} (full line), which do not include radiative corrections. It is seen that 
the two calculations agree well, which implies that the effect of radiative corrections is very small compared to the experimental errors. 

\begin{figure}
\begin{center}
\epsfig{file=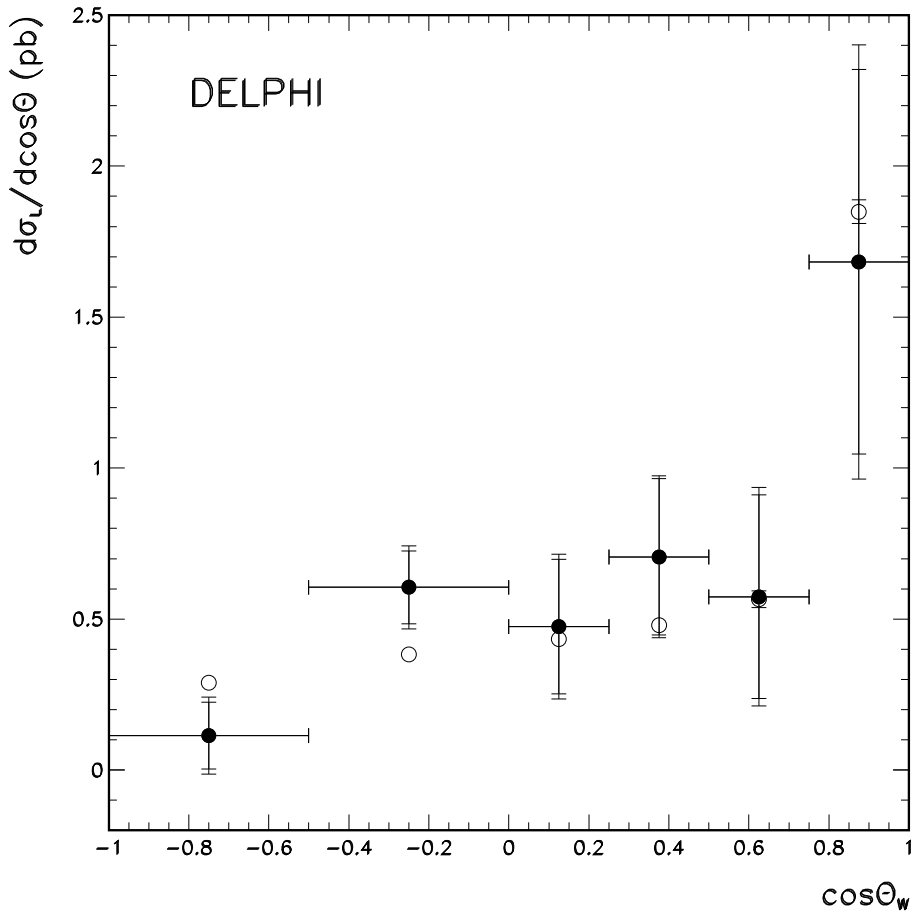 ,width=14cm}
\caption[]{Luminosity weighted average of the differential cross-sections measured at 189, 198 and 206 GeV (black dots) for longitudinally polarised W-bosons as a function of 
$\cos \Theta_W$, with statistical and total errors. The open circles show the values obtained from WPHACT MC at 199.5 GeV at generator level.
}
\label{fig:long-Xsec}
\end{center}
\end{figure}

The differential cross-section  for the production of longitudinally polarised W bosons is 
\begin{eqnarray}
\label{equ:polaL}
\frac{d\sigma_L}{d\cos\Theta_W}& = &\rho^W_{00}(\cos\Theta_W)
                                 \frac{d\sigma}{d\cos\Theta_W}. 
\label{equ:polaT}
\end{eqnarray}
In this formula 
$d\sigma/d\cos\Theta_W$ is the differential cross-section after correction for detector acceptance and sample purity.
The differential cross-sections were determined for the three energies considered.
Figure \ref{fig:long-Xsec} shows the luminosity weighted average of the 
 measured differential cross-sections, together with the Standard Model predictions from WPHACT. 
The two distributions are in good agreement.

\begin{figure}
\begin{center}
\epsfig{file=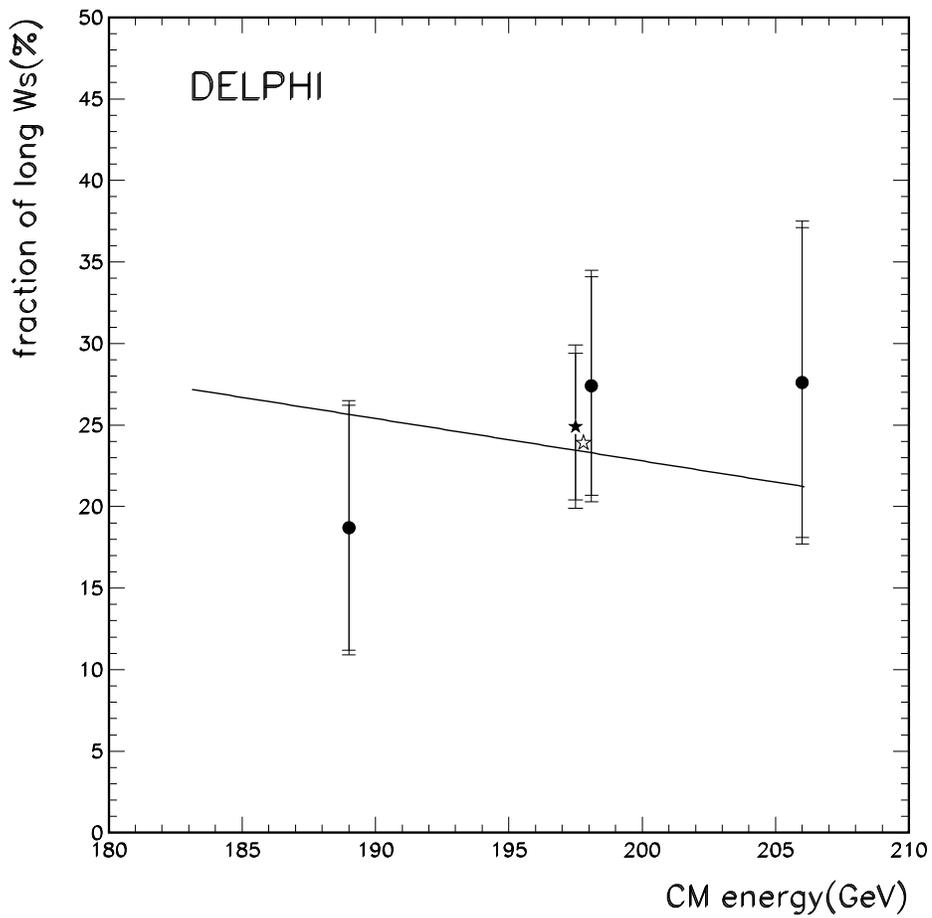 ,width=14cm}
\caption[]{Fraction of longitudinally polarised W-bosons as function of centre-of-mass energy, with statistical and total errors. The black dots represent the measurements and the 
full line the values obtained from WPHACT MC at generator level. 
The black star is the luminosity weighted mean of the measurements at the three energies and the open star the equivalent mean obtained from WPHACT MC at generator level as explained in the text. 
}
\label{fig:fl}
\end{center}
\end{figure}

Integration  yields the fraction of longitudinally polarised W bosons: 
\begin{eqnarray}
f_L = \sigma_L/\sigma_{tot}.
\end{eqnarray}
Values of 
$18.7 \pm 7.5$ \%, $ 27.4 \pm 6.7$ \% and $27.6 \pm 9.5$ \%
are obtained from the data at 189, 198 and 206 GeV respectively, while values of 
$ 25.8 \pm 0.3$  \%, $23.4 \pm 0.3$ \% and $22.6 \pm 0.3$ \%
are expected from the Standard Model Monte Carlo (about 50000 $\rm pb^{-1}$ at each energy). 
These errors are statistical only.  
The fraction of longitudinal W bosons is shown as a function of the energy in figure \ref{fig:fl}. 
The luminosity weighted average over the three data samples is 
\begin{eqnarray}
\sigma_L/\sigma_{tot} = 
24.9 \pm 4.5 (stat)\pm 2.2 (syst) \% 
\end{eqnarray}
at a mean energy of 198 GeV. 
The systematic error is discussed in section \ref{sec:systematics}. 
This is in good agreement with 
the corresponding value 
of $ 23.9 \pm 0.2$ \% expected from Standard Model Monte Carlo.

\section{Systematic errors on the SDM elements}
\label{sec:systematics}

The systematic uncertainties in the measurements of the SDM elements were calculated as described below. 
The list of systematic errors considered for $\rho_{00}$ 
is shown in table \ref{tab:r00-systematics-200} as an example. 
The systematic errors on the differential cross-section  and on the fraction of longitudinally polarised W bosons were estimated in the same way and are discussed at the end of this section.

\begin{enumerate}
\item 
Monte Carlo statistics. The detector corrections are binned in 8 bins in $\cos \Theta_W$, 10 bins in $\cos\theta^*$ and 10 bins in $\phi^*$. Some bins have a low population of events which results in a large uncertainty in the correction factor. To estimate this effect on the SDM elements, the simulated data samples were divided in 9 subsamples of about 2600 $\rm pb^{-1}$ and detector corrections were computed for each subsample. The analysis was rerun on the data with each set of detector corrections and the differences of the new SDM elements with the SDM elements obtained with the standard corrections were computed. The standard deviation of the distributions of differences, corrected for the factor 9 difference in statistics between the subsamples and the full sample,  was taken as the systematic error. 
 
 \item
Signal and background cross-sections.  The uncertainties on the signal and background cross-sections influence the purities. For the estimation of the systematic error arising from the uncertainty on the background cross-sections only the uncertainties on the $q \bar{q} (\gamma)$ and four-fermion neutral current cross-sections were taken into account, and were taken to be 5\% 
\cite{ref:yellow-precision}.
 The purities were recalculated with  background cross-sections which were modified by plus and minus one standard deviation. The mean of the differences of the recomputed SDM elements  and the standard elements was taken as systematic uncertainty. \\
The uncertainty on the signal cross-section enters both in the denominator and the numerator and its effect is expected to be small.  
The purities were recalculated with signal cross-sections which were modified by plus and minus one standard deviation. 
The uncertainty on the signal cross-section  was taken to be 0.5\% , the theoretical error 
\cite{ref:yellow-precision}.
The mean of the differences of the recomputed SDM elements  and the standard elements was taken as systematic uncertainty. These uncertainties are negligible at all energies considered.
\item 
Jet reconstruction, hadronisation modelling and migration of events between $\cos \Theta_W$ bins. The reconstruction of the hadronic jets influences the determination of the W production and decay angles and will hence lead to migration effects between bins in the $\cos \Theta_W$ distribution. 
On the other hand, the corrections for acceptance and purity are sensitive to the modelling of the hadronisation in the simulation. 
To estimate these effects, 
the differences between the SDM elements calculated with simulated events at generator level and at reconstruction level, using the HERWIG hadronisation modelling \cite{ref:HERWIG}, were computed. 
The reconstructed SDM elements were obtained by reweighting the selected events with the standard detector corrections obtained from the JETSET hadronisation modelling.
The absolute values of these differences were taken as systematic uncertainty. This uncertainty was estimated at 199.5~GeV and the same value was used for all 3 energies.
A problem with the track reconstruction efficiency for low-momentum particles at low polar angles was corrected for as described in \cite{ref:delphi-cms-energy}. We have investigated the systematic error related to this correction and found that it was negligible.

\item 
Cut on lepton polar angle. 
In the analysis, events with a lepton close to the beam (polar angle below $20 ^{\circ}$ or above $160 ^{\circ}$) were rejected, and the standard detector corrections were calculated accordingly. 
To estimate the effect of the limited resolution in the reconstruction of the lepton angle, the analysis was redone with a cut at both $18 ^{\circ}$ and $22 ^{\circ}$. 
The detector corrections were recalculated, one set for each cut, and the events were corrected with these new sets. 
The differences between the SDM elements obtained in the analysis with a cut at $22  ^{\circ}$  and the  analysis with a cut at $18 ^{\circ}$ were rescaled to a difference 
corresponding to $\pm 0.5^{\circ}$.
This is a conservative estimate compared to the estimated value of the resolution which is about $0.1^{\circ}$, plus some tails.
In addition, the SDM elements were recalculated with these new cuts, but corrected with the standard detector corrections, and the difference scaled down to 
$\pm 0.5^{\circ}$ was also computed. 
This yields two estimates of the uncertainty related to the resolution on the lepton polar angle reconstruction and the modelling of this reconstruction in the simulation. 
The larger estimate was taken as systematic uncertainty.

\item 
Cut on the $\chi^2$ probability of the 3C fit. The analysis was redone with two different cuts on the $\chi^2$ probability, at 0.003 and at 0.007, in a region where the probability has a flat distribution. For each cut, detector corrections were recalculated and the data were corrected with these new sets of corrections. The mean difference between the elements obtained with each new set of corrections and the standard elements was taken as systematic uncertainty.

\item 
Radiative corrections and CC03 reweighting. 
The purities which enter in the detector corrections refer to CC03 events of the type $e^+ e^- \rightarrow W^+ W^- \rightarrow l \nu q \bar{q} (l=e,\mu)$. 
The simulated event samples which were used to calculate these purities contain all four-fermion charged current processes. 
To obtain the signal angular distributions which are input to the purity calculations the events were reweighted with CC03 weights following the reweighting procedure explained in ref. \cite{ref:dpaweighting}. 
The uncertainty on the calculation of the radiative corrections has only a small influence on the SDM elements (see section \ref{sec:sdm}). 
The combined effect of the uncertainty from the CC03 reweighting and the radiative corrections was estimated by the difference between the analytical calculation of the SDM's used for the TGC fits (CC03 in the zero width approximation, no radiative corrections at all, see 
\cite{ref:gounaris})
and the SDM elements calculated at generator level with samples of simulated signal events corresponding to about 50000 $\rm pb^{-1}$ 
(WPHACT MC).
For the cases where the error on the Monte Carlo calculation was larger than this difference, this error was taken as systematic uncertainty.

\item 
Lepton charge determination. In the forward and backward regions of the detector the lepton charge is sometimes badly determined. To estimate this effect on the SDM elements, 10\% of the events
were artificially given a wrong charge and the elements were recalculated with standard detector corrections. From a study of two-lepton events \cite{ref:two-fermion} the fraction of leptons with a wrong charge assignment was estimated to be less than 1\%. 
The uncertainty on the SDM elements from lepton charge determination was obtained from a rescaling by a factor 10 of the difference between the elements calculated with the 10\% wrong charge data and the standard elements.

\end{enumerate}
The systematic errors on the 9 SDM elements in a given bin at a given energy are fully correlated since 
the elements are determined from the same events. 
The systematic error from Monte Carlo statistics (1.) is uncorrelated between bins and energies. 
All other systematic errors are fully correlated between bins and energies.
Therefore 
a luminosity weighted average of the values obtained at the three energies was used
in the TGC fits, hence reducing the effects of statistical fluctuations.

The systematic errors on the differential cross-sections and the fraction of longitudinally polarised 
W bosons 
were estimated with the same procedure as that used for the SDM elements. 
When computing the luminosity weighted average of these quantities all systematic errors were considered fully correlated between years, apart from the error from Monte Carlo statistics.
The systematic error on the fraction $f_L$ is given in table \ref{tab:fl-systematics}.

\section{Fits of Triple Gauge Couplings}
\label{sec:tgc}
Both CP-conserving and CP-violating TGC's 
are determined in this analysis, which is however particularly suited to the determination of CP-violating couplings, whose existence would 
be revealed by non-zero imaginary parts of the SDM's.
To investigate the possible existence of the anomalous 
CP-violating TGC's  $g_4^Z, \tilde{\kappa}_Z, \tilde{\lambda}_Z$ in each of the three data samples defined in table \ref{tab:statistics}, the experimental values of the single W SDM elements $\rho_{\tau \tau'}^{W^-}(s,\cos\Theta_W)$ and $\rho_{\tau \tau'}^{W^+}(s,\cos\Theta_W)$ determined in each of the 
$\cos\Theta_W$ bins considered in this analysis 
were fitted to theoretical expressions derived in Ref. \cite{ref:gounaris}. 
For CP-invariant interactions the relationship 
(\ref{equ:rho-wmwp}) holds. 
This allows a combination of $\rm W^-$ and $\rm W^+$ elements in each $ \cos\Theta_W$ bin. 
This procedure was applied in order to extract the CP-conserving couplings $ \Delta g_1^Z, \Delta \kappa_{\gamma}$ and $\lambda_{\gamma}$.

In each of the $\cos\Theta_W$ bins the 9 SDM elements are correlated. 
The strongest correlations occur between  $\rho_{++},\rho_{--}$ and $ \rho_{00}$, whose sum is constrained to be one. 
The correlations were determined from the data and taken into account in the fit.

As the sum of the projection operators $\Lambda_{++}+\Lambda_{--}+\Lambda_{00} = 1$, it is seen from expression (\ref{equ:rho-exp}) that the sum of the experimentally determined diagonal SDM elements will always be exactly equal to one, whatever the sample used. 
The most straightforward way to take this constraint into account is to retain only two of the three diagonal elements in the fit, whose results are indeed totally insensitive to which of those elements is rejected. 
In the following, the element $\rho_{++}$ has been removed from the fits which 
are hence reduced to five real SDM elements per bin ($\rho_{--},\rho_{00},Re(\rho_{+-}),Re(\rho_{+0}),Re(\rho_{-0})$) to determine the CP-conserving couplings, and to sets of 8 elements per bin (as above plus $Im(\rho_{+-}),Im(\rho_{+0}),Im(\rho_{-0})$) for the extraction of the CP-violating couplings.

A least squares fit was used in which the measured values of the SDM elements were compared to their theoretical predictions 
at the average centre-of-mass energies for each of the three 
data sets. 
The statistical covariance matrices were computed from the data.
These were combined with the full systematic covariance matrix containing the systematic errors described in section \ref{sec:systematics}. 

\begin{figure}
\epsfig{file=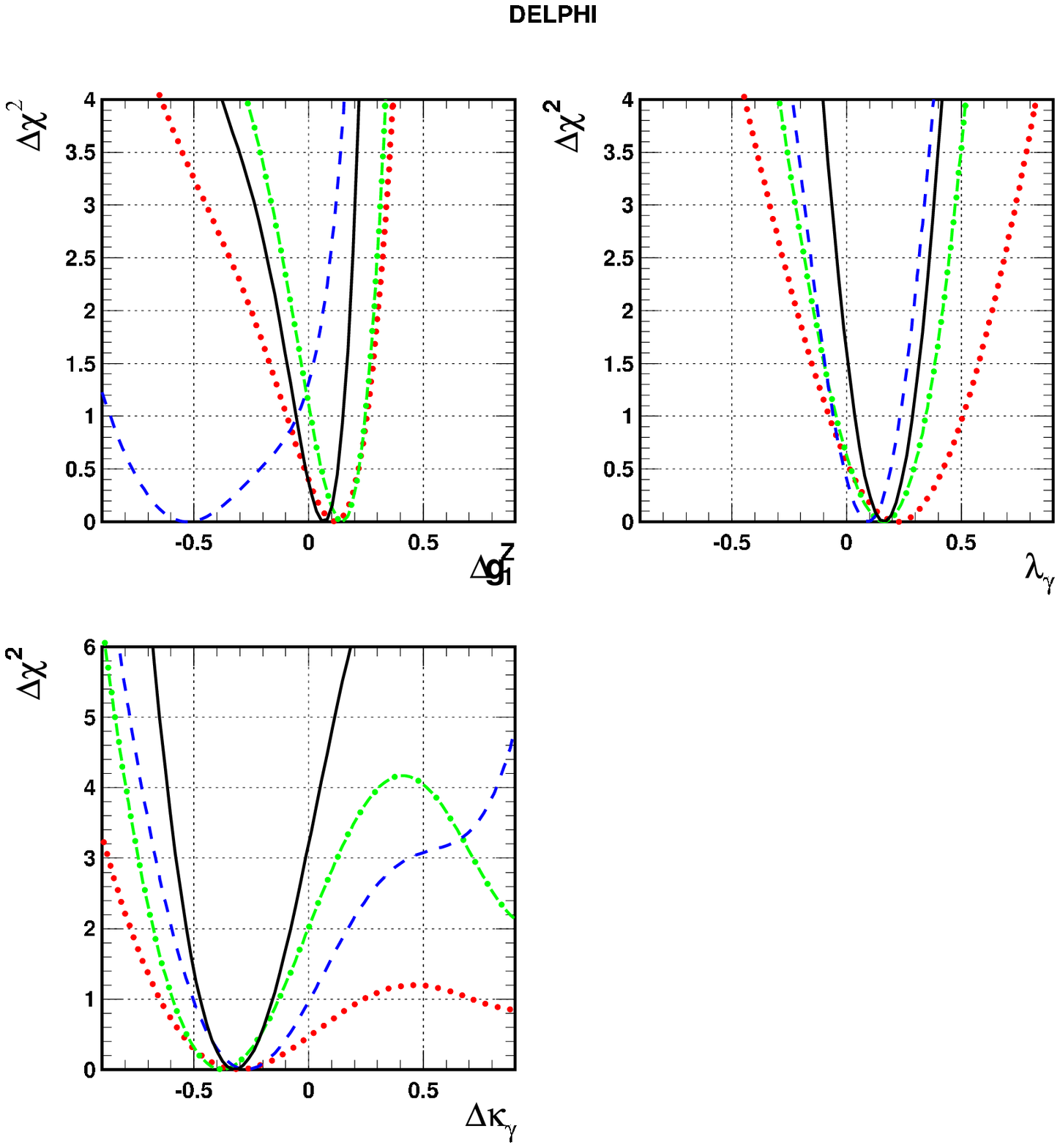,width=14cm}
\caption[]{
Results of the one-parameter CP-conserving TGC fits. 
The full lines show the $\chi^2$ curves for the full data sample, the dotted lines show the 189 GeV results, the dash-dotted lines show the results at 198 GeV and the dashed lines show the results at 206 GeV. Statistical and systematic errors are included. The results of the fits are displayed in table \ref{tab:one-para-syst}. }
\label{fig:chi2-1para-CPeven}
\end{figure}

\begin{figure}
\epsfig{file=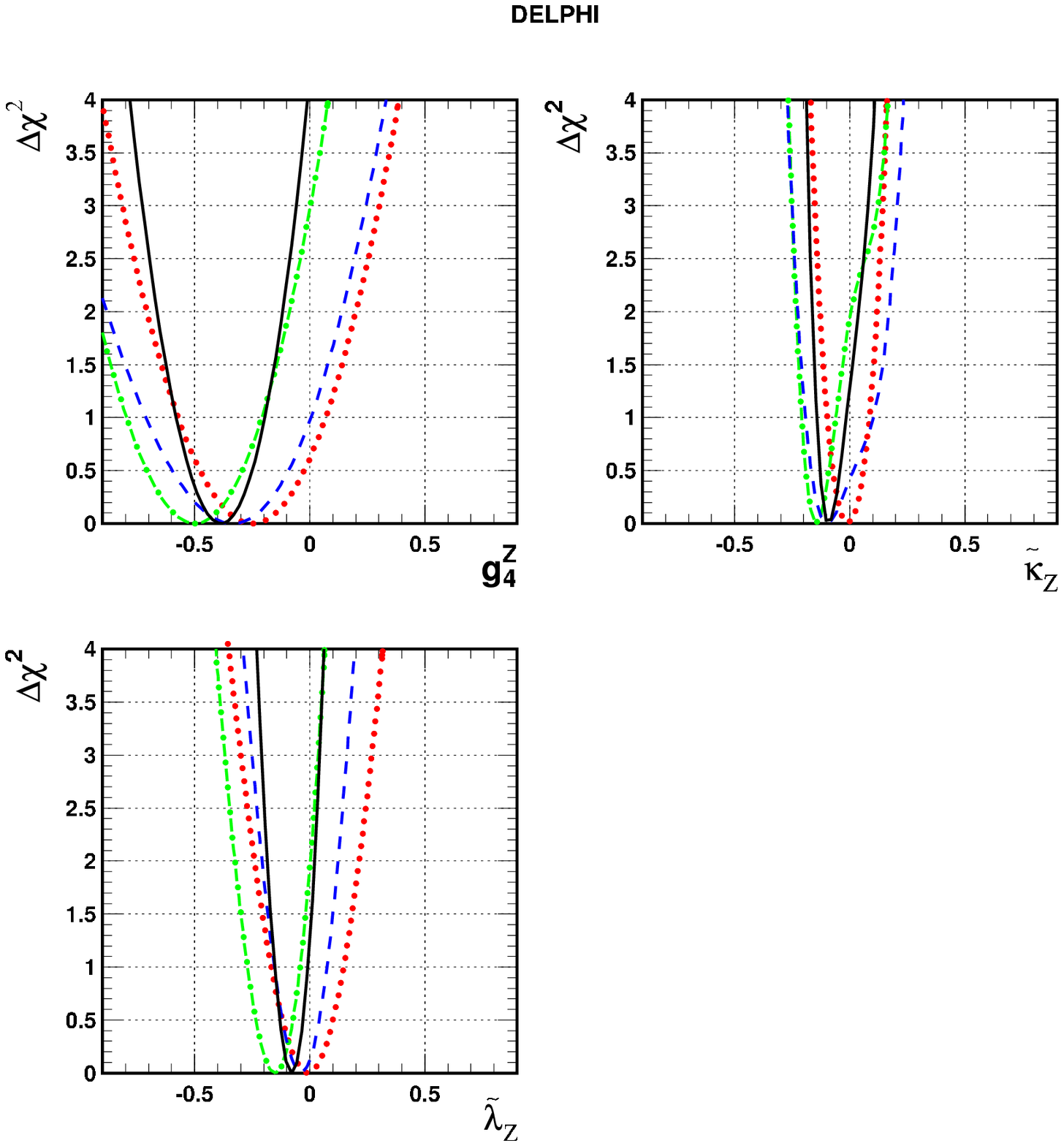,width=14cm}
\caption[]{
Results of the one-parameter CP-violating TGC fits. 
The full lines show the $\chi^2$ curves  for the full data sample, the dotted lines show the 189 GeV results, the dash-dotted lines show the results at 198 GeV and the dashed lines show the results at 206 GeV. Statistical and systematic errors are included. The results of the fits are displayed in table \ref{tab:one-para-syst}. }
\label{fig:chi2-1para-CPodd}
\end{figure}

Table \ref{tab:one-para-syst}
shows the results of the one-parameter fits for the three data sets separately and for the combined fit to all data. 
The total (statistical and systematic) error matrices were used.
In each $\chi^2$ fit only one of the TGC's considered was varied, all other couplings being fixed at their SM value. 
The $\chi^2$ curves of the fits 
are displayed in figure \ref{fig:chi2-1para-CPeven} for the CP-conserving couplings and in figure \ref{fig:chi2-1para-CPodd} for the CP-violating couplings. 
The minimum $\chi^2$ values are displayed in  table \ref{tab:one-para-syst}.
The $\chi^2$ probabilities of all fits to the full sample are acceptable, but are considerably lower for the CP-violating fits than for the CP-conserving fits. 
This is mainly due to the data at 189 GeV. 
The errors on the results of fits using only statistical errors on the SDM elements are given in the last column of table \ref{tab:one-para-syst}.
It is seen that the results of the fits are dominated by the statistical errors.
Using statistical errors only, the results of the Monte Carlo studies of 250 data-sized samples with SDM's computed at generation and at reconstruction level do not indicate any marked bias of the fitted values of the TGC's with respect to their SM input values. 
These Monte Carlo studies also revealed the existence of a double minimum in the fits of $\Delta \kappa_{\gamma}$ which is confirmed by the data, as seen in figure \ref{fig:chi2-1para-CPeven}. Such double minima can occur \cite{ref:yellow-report-TGC,ref:sekulin} as the helicity amplitudes are linear in the couplings. 

In the fits to the data the average beam energies, displayed in table \ref{tab:statistics} for each of the data taking years, were used. 
However, as already mentioned in section \ref{sec:datasample}, the beam energy of the data samples taken in 1999 varied from 192 to 202 GeV and from 204 to 209 GeV for the samples taken in the year 2000. 
The effect of these beam energy spreads on the errors on the fitted values of the TGC's was estimated by repeating the single parameter fits with beam energy values varying within the allowed energy ranges. 
The resulting shifts in the fitted values of the TGC parameters are very small and have been treated as systematic errors included in the full errors given in table \ref{tab:one-para-syst}. 
The maximum size of this systematic error is 0.02. 

\begin{figure}
\epsfig{file=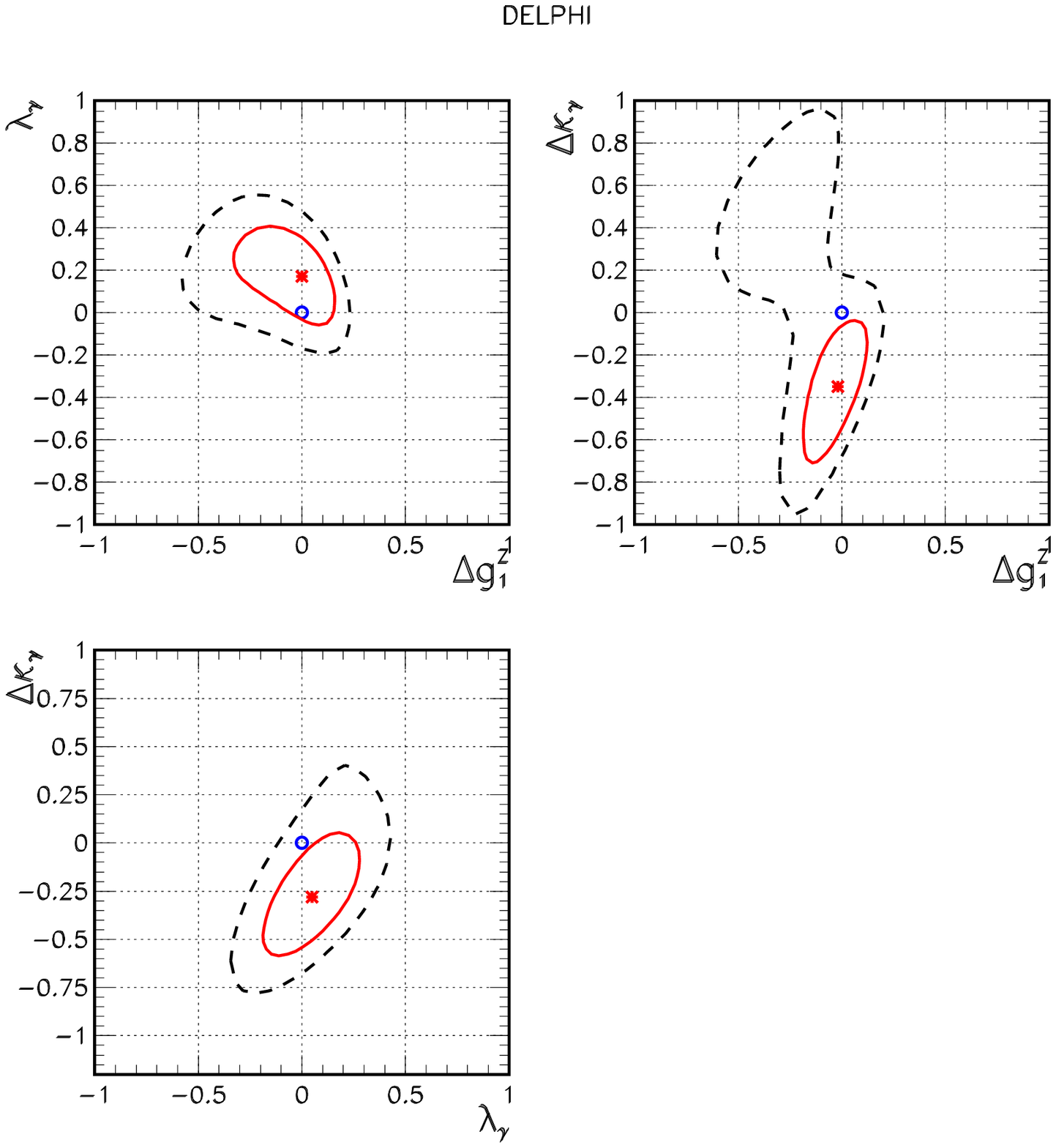,width=14cm}
\caption[]{Two-parameter CP-conserving TGC fits to the full data set. The star shows the fit results while the open circle represents the SM value. The full line shows the 68\% CL contour and the dashed line the 95\% CL contour. Statistical and systematic errors are included. }
\label{fig:two-para-CPeven}
\end{figure}

\begin{figure}
\epsfig{file=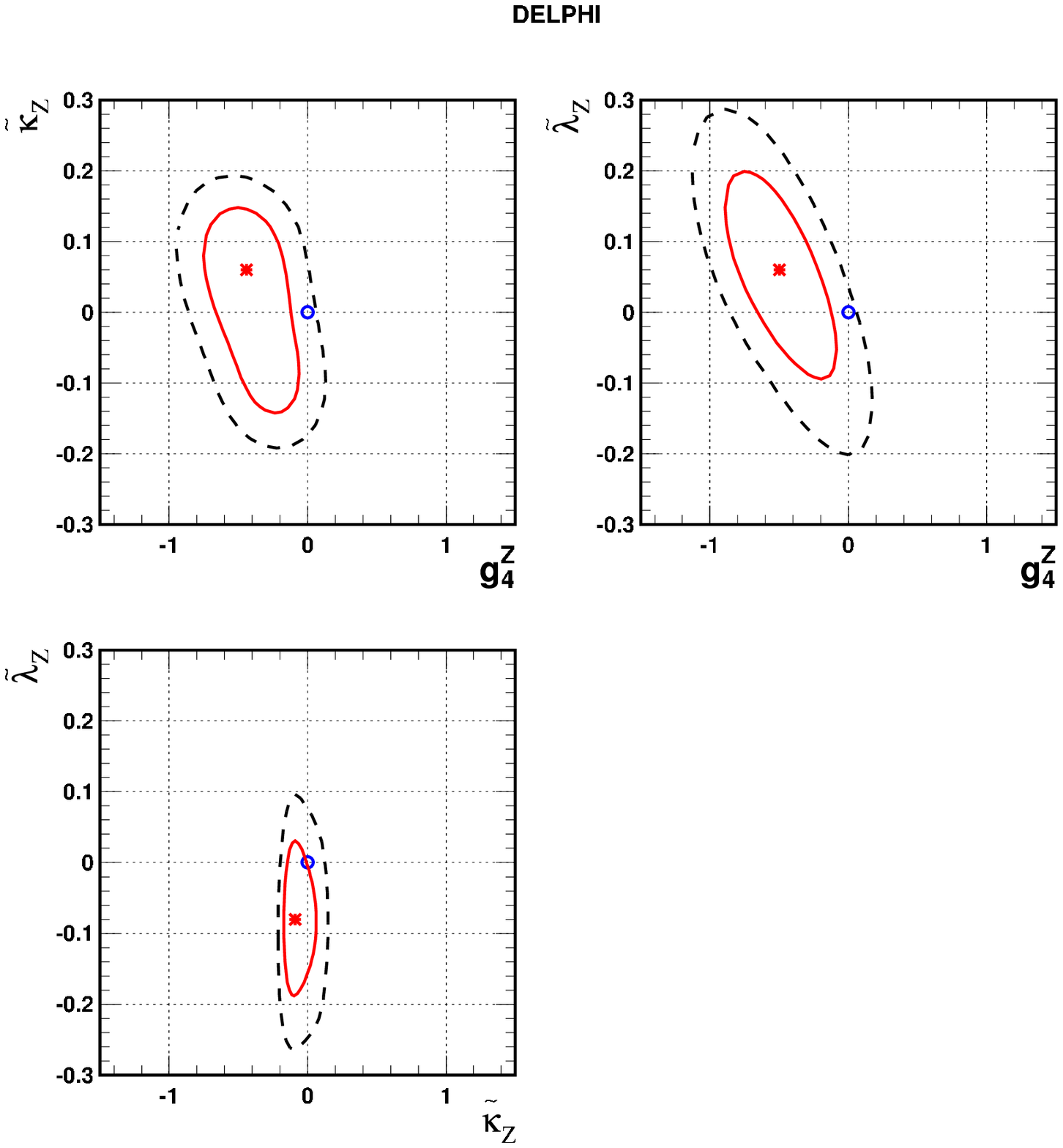,width=14cm}
\caption[]{Two-parameter CP-violating TGC fits to the full data set. The star shows the fit results while the open circle represents the SM value. The full line shows the 68\% CL contour and the dashed line the 95\% CL contour. Statistical and systematic errors are included. }
\label{fig:two-para-CPodd}
\end{figure}

Two-parameter fits of the TGC's  at fixed central beam energy values were also performed, the results of which are shown in figures \ref{fig:two-para-CPeven} and \ref{fig:two-para-CPodd} for the full data set using the total (statistical and systematic) error matrix.
The results are in reasonable agreement with the SM expectations. 
It is seen from figure \ref{fig:two-para-CPeven} that the fit of $\Delta \kappa_{\gamma}$ exhibits a second minimum which appears as an extension of the 95\% probability contour. 
This second minimum also 
strongly affects the shape of the $\Delta \chi^2$-plot at 189 GeV shown in figure \ref{fig:chi2-1para-CPeven}.

Finally, three-parameter fits to the full 
data sample with full error matrices were also performed separately for the CP-conserving and CP-violating couplings respectively. 
The results are shown in table \ref{tab:three-para}, in which the errors shown are the standard deviations of the marginal distributions of each of the parameters.

The results of the one, two and three-parameters fits are consistent with each other and agree with the Standard Model.
\section{Summary}
\label{sec:summary}
The data taken by the DELPHI experiment at centre-of-mass energies of 189, 192-202 and 204-209 GeV
were used to select a sample of respectively 520, 838 and 522 events of the type 
$ e^+ e^- \rightarrow l \nu q \bar{q}(l=e,\mu)$. 
The decay angles of the
leptonically decaying 
W bosons 
were used to calculate the single $\rm W^-$ and $\rm W^+$ spin density matrices, 
which are defined for CC03 events, 
and the average values assuming CP symmetry.

The SDM elements were used to determine the fractions of longitudinally polarised 
W bosons. 
For each of the three data samples the measured  fraction of longitudinally polarised W bosons is in agreement with the SM prediction.
For all data taken between 189 and 209 GeV an average value of 
\begin{eqnarray}
\sigma_L/\sigma_{tot} = 
24.9 \pm 4.5(stat) \pm 2.2(syst) \%
\end{eqnarray}
is obtained at an average energy of 198 GeV, where 
$23.9 \pm 0.2 \%$ is expected from the Standard Model.

The SDM elements have been used to determine the CP-violating Triple Gauge Couplings. 
One-parameter fits to the full data sample yield:
\begin{eqnarray*}
g_4^Z             & = &-0.39 ^{+0.19}_{-0.20} \\
\tilde{\kappa}_Z  & = &-0.09 ^{+0.08}_{-0.05} \\
\tilde{\lambda}_Z & = &-0.08 \pm 0.07.
\end{eqnarray*}

\noindent
For the CP-conserving TGC's the results are:
\begin{eqnarray*}
\Delta g_1^Z          & = &0.07^{+0.08}_{-0.12} \\
\lambda_{\gamma}      & = &0.16^{+0.12}_{-0.13}  \\
\Delta \kappa_{\gamma}& = &-0.32^{+0.17}_{-0.15}.
\end{eqnarray*}

\noindent
The errors quoted result from a quadratic combination of the statistical and systematic errors on the SDM elements.

For the CP-conserving TGC's the values obtained in this analysis are less precise than those measured in the DELPHI analysis using optimal observables \cite{ref:DELPHI-tgc}, but they confirm the good agreement of all the fitted couplings with the predictions of the Standard Model.



\subsection*{Acknowledgements}
\vskip 3 mm
We are greatly indebted to our technical 
collaborators, to the members of the CERN-SL Division for the excellent 
performance of the LEP collider, and to the funding agencies for their
support in building and operating the DELPHI detector.\\
We acknowledge in particular the support of \\
Austrian Federal Ministry of Education, Science and Culture,
GZ 616.364/2-III/2a/98, \\
FNRS--FWO, Flanders Institute to encourage scientific and technological 
research in the industry (IWT) and Belgian Federal Office for Scientific, 
Technical and Cultural affairs (OSTC), Belgium, \\
FINEP, CNPq, CAPES, FUJB and FAPERJ, Brazil, \\
Ministry of Education of the Czech Republic, project LC527, \\
Academy of Sciences of the Czech Republic, project AV0Z10100502, \\
Commission of the European Communities (DG XII), \\
Direction des Sciences de la Mati$\grave{\mbox{\rm e}}$re, CEA, France, \\
Bundesministerium f$\ddot{\mbox{\rm u}}$r Bildung, Wissenschaft, Forschung 
und Technologie, Germany,\\
General Secretariat for Research and Technology, Greece, \\
National Science Foundation (NWO) and Foundation for Research on Matter (FOM),
The Netherlands, \\
Norwegian Research Council,  \\
State Committee for Scientific Research, Poland, SPUB-M/CERN/PO3/DZ296/2000,
SPUB-M/CERN/PO3/DZ297/2000, 2P03B 104 19 and 2P03B 69 23(2002-2004),\\
FCT - Funda\c{c}\~ao para a Ci\^encia e Tecnologia, Portugal, \\
Vedecka grantova agentura MS SR, Slovakia, Nr. 95/5195/134, \\
Ministry of Science and Technology of the Republic of Slovenia, \\
CICYT, Spain, AEN99-0950 and AEN99-0761,  \\
The Swedish Research Council,      \\
Particle Physics and Astronomy Research Council, UK, \\
Department of Energy, USA, DE-FG02-01ER41155, \\
EEC RTN contract HPRN-CT-00292-2002. \\

We also want to thank J. Layssac for useful comments on the interpretation of the theoretical expressions.


\newpage


\newpage

\begin{table}
\begin{center}
\begin{tabular}{|l|ccc|}  \hline
data taking year                  & 1998              &  1999             &  2000               \\  \hline 
mean energy (GeV)         & 189               &  198              &  206               \\
energy range (GeV)                & [188.5 - 189.]    &  [191.5 - 202.]   &  [204. - 209.]      \\
luminosity ($\rm pb^{-1}$)        & 153.8             &  218.0            &  148.6              \\
e+$\rm \mu$ after all cuts (\# evts)  & 520               &  838              &  522                \\ \hline
efficiency electron evts              &     0.656         &     0.639         &     0.628           \\
efficiency muon evts                  &     0.787         &     0.759         &     0.743           \\
average efficiency e+$\rm \mu$        &     0.721         &     0.699         &     0.685           \\
average purity e+$\rm \mu$            &     0.923         &     0.917         &     0.914           \\ \hline
energy of MC sample (GeV)            & 188.6             &  199.5            &  206               \\
MC statistics CC ($\rm pb^{-1}$)      & 26600             & 25000             & 24600               \\
MC statistics NC   ($\rm pb^{-1}$)    & 18400             & 10000		  & 19000               \\
MC statistics $q \bar{q} (\gamma)$ ($\rm pb^{-1}$)& 5000  & 5700              & 6300                \\ \hline 
\end{tabular}
\end{center}
\caption[]{Statistics collected in each data taking year, Monte Carlo (MC) statistics used to calculate the detector corrections, efficiencies and purities. The Monte Carlo  simulations have been performed at fixed centre-of-mass energies, as discussed in the text.}
\label{tab:statistics}
\end{table}

\newpage

\begin{table}
\begin{center}
\begin{tabular}{|c|cccccc|} \hline
$\cos \Theta_W$ bin & 1 & 2 & 3 & 4 & 5 & 6 \\ \hline
MC statistics             & 0.042 &  0.029 &  0.021 &  0.011 &  0.017 &  0.008\\
theoretical cross-sections& 0.003 &  0.001 &  0.001 &  0.002 &  0.001 &  0.001  \\
reconstruction            & 0.006 &  0.012 &  0.034 &  0.020 &  0.003 &  0.027   \\
$\theta_{lepton}$ cut     & 0.026 &  0.007 &  0.005 &  0.009 &  0.016 &  0.017    \\
$\rm Prob(\chi^2)$ cut    & 0.021 &  0.023 &  0.027 &  0.007 &  0.017 &  0.008 \\
radiat. corr. + CC03 rewgt& 0.019 &  0.010 &  0.010 &  0.009 &  0.014 &  0.032  \\
lepton charge             & 0.018 &  0.010 &  0.005 &  0.006 &  0.004 &  0.003    \\ \hline
total systematic error    & 0.060 &  0.042 &  0.050 &  0.028 &  0.033 &  0.047  \\ \hline

\end{tabular}
\caption{Luminosity weighted average of the systematic error on $\rm \rho_{00}$ (average of $\rm W^-$ and $\rm W^+$ elements) in the 6 $\cos\Theta_W$ bins  with bin 1 being the most backward bin.  }
\label{tab:r00-systematics-200}
\end{center}
\end{table}


\begin{table}
\begin{center}
{
\begin{tabular}{|c|cccc|} \hline
                  Data set&  189 GeV&  198 GeV&   206 GeV& average \\ \hline
MC statistics             &    0.010&    0.011&    0.014 & 0.007\\
theoretical cross-sections&    0.001&    0.002&    0.002 & 0.002\\
reconstruction            &    0.015&    0.015&    0.015 & 0.015\\
$\theta_{lepton}$ cut     &    0.007&    0.007&    0.011 & 0.008\\
$\rm Prob(\chi^2)$ cut        &    0.004&    0.004&    0.005 & 0.005\\
radiat. corr. + CC03 rewgt&    0.007&    0.010&    0.016 & 0.011\\
lepton charge             &    0.002&    0.001&    0.003 & 0.002 \\ \hline
total systematic error    &    0.021&    0.023&    0.029 & 0.022\\
statistical error         &    0.075&    0.067&    0.095 & 0.045\\ \hline
\end{tabular}
\caption{Systematic error on $ f_L$ for the 3 energies and luminosity weighted average.}
\label{tab:fl-systematics}
}
\end{center}
\end{table}


\begin{table}
\begin{center}
{
\begin{tabular}{|crrr|cr|} \hline
{} &{} &{} &{} &{} &{}   \\
Data set & 189 GeV & 198 GeV & 206 GeV & \multicolumn{2}{|c|}{full sample}  \\
        &         &         &         & fit result    & stat. err.\\ 
{} &{} &{} &{} &{} &{}   \\ \hline
{} &{} &{} &{} &{} &{}   \\ 
$\Delta g_1^Z$          &0.12 $^{+0.14}_{-0.21}$&0.15$^{+0.10}_{-0.14}$&-0.53$^{+0.48}_{-0.34}$&0.07$^{+0.08}_{-0.12}$ &$^{+0.08}_{-0.10}$\\
{} &{} &{} &{} &{} &{}   \\ 
 $\rm \chi^2/ndf$        & 23/29    & 18/29    & 18/29   & 61/89  &   \\\hline
{} &{} &{} &{} &{} &{}   \\ 
$\lambda_{\gamma}$      &0.22 $^{+0.29}_{-0.30}$&0.16$^{+.18}_{-.21}$  &0.09$^{+0.14}_{-0.15}$ &0.16$^{+0.12}_{-0.13}$ &$^{+0.08}_{-0.09}$\\
{} &{} &{} &{} &{} &{}   \\ 
 $\rm \chi^2/ndf$        & 23/29    & 19/29    & 19/29   &  60/89 &   \\\hline
{} &{} &{} &{} &{} &{}   \\ 
$\Delta \kappa_{\gamma}$&-0.31$^{+0.55}_{-0.34}$&-0.38$^{+.25}_{-.22}$ &-0.27$^{+0.28}_{-0.23}$&-0.32$^{+0.17}_{-0.15}$&$^{+0.16}_{-0.14}$\\ 
{} &{} &{} &{} &{} &{}   \\ 
 $\rm \chi^2/ndf$        & 23/29    & 17/29    & 18/29   & 58/89   &   \\
\hline \hline
{} &{} &{} &{} &{} &{}   \\ 
$g_4^Z$            & -0.25$\pm$ 0.32 & -0.50$^{+0.29}_{-0.30}$ &-0.34$^{+0.34}_{-0.38}$ &-0.39$^{+0.19}_{-0.20}$&$\pm0.17$\\
{} &{} &{} &{} &{} &{}   \\ 
 $\rm \chi^2/ndf$        & 141/95    & 108/95    & 80/95     & 330/287 &   \\\hline
{} &{} &{} &{} &{} &{}   \\ 
$\tilde{\kappa}_Z$ & -0.01$\pm$ 0.09 & -0.15$^{+0.09}_{-0.06}$ &-0.10$^{+0.20}_{-0.09}$ &-0.09$^{+0.08}_{-0.05}$&$^{+0.06}_{-0.05}$\\
{} &{} &{} &{} &{} &{}   \\ 
 $\rm \chi^2/ndf$        & 142/95    & 109/95    & 81/95     & 333/287 &   \\\hline
{} &{} &{} &{} &{} &{}   \\ 
$\tilde{\lambda}_Z$& -0.01$\pm$ 0.16 & -0.15$^{+0.11}_{-0.12}$ &-0.04$^{+0.11}_{-0.12}$ &-0.08$\pm$0.07         &$\pm0.07$\\
{} &{} &{} &{} &{} &{}   \\ 
 $\rm \chi^2/ndf$        & 142/95    & 109/95    & 81/95    &333/287  &   \\
\hline
\end{tabular}
\caption{ Results of one-parameter fits including total (statistical and systematic) errors.
In the last column, the errors on the results of fits to the full sample using only statistical errors on the SDM elements are given for comparison.}
\label{tab:one-para-syst} 
}
\vspace*{-13pt}
\end{center}
\end{table}



\begin{table}
\begin{center}
{
\begin{tabular}{@{}crrrr@{}}
\hline
{} &{} &{} &{} &{}\\[-1.5ex]
 & fitted value & $\Delta g_1^Z$  & $\lambda_{\gamma}$  & $\Delta \kappa_{\gamma}$ \\[1ex]
\hline
{} &{} &{} &{} &{}\\[-1.5ex]
$\Delta g_1^Z$          &-0.03 $^{+0.10}_{-0.11}$ & 1.00 & -0.22 & 0.47 \\[1ex] 
$\lambda_{\gamma}$      & 0.06 $\pm0.16$          &      & 1.00  & 0.45 \\[1ex] 
$\Delta \kappa_{\gamma}$&-0.31 $^{+0.24}_{-0.25}$ &      &       & 1.00 \\[1ex] 
\hline
\hline
{} &{} &{} &{} &{}\\[-1.5ex]
 & fitted value & $g_4^Z$ & $\tilde{\kappa}_Z$ & $\tilde{\lambda}_Z$  \\[1ex]
\hline
{} &{} &{} &{} &{}\\[-1.5ex]
$g_4^Z$            & -0.58 $\pm0.27$         & 1.00     & -0.23   & -0.66 \\[1ex]
$\tilde{\kappa}_Z$ &  0.06 $^{+0.07}_{-0.10}$&          & 1.00   & 0.06  \\[1ex]
$\tilde{\lambda}_Z$&  0.07 $\pm$0.09         &          &        & 1.00 \\[1ex]
\hline
\end{tabular}
\caption{ Results of three-parameter fits to the full sample. The errors are the total, statistical plus systematic, uncertainties. The $\chi^2$ for the fits of the CP-conserving parameters (top) is 58 for 87 degrees of freedom. The $\chi^2$ for the fits of the CP-violating parameters (bottom) is 329 for 285 degrees of freedom. }
\label{tab:three-para} 
}
\vspace*{-13pt}
\end{center}
\end{table}











\newpage


\clearpage 

\end{document}